\documentclass[12pt,letterpaper]{article}
\usepackage{etoolbox}
\newtoggle{SUPPLEMENTAL}\toggletrue{SUPPLEMENTAL}
\togglefalse{SUPPLEMENTAL} %COMMENT OUT FOR SUPPLEMENTAL APPENDIX
\newcommand{\Supplemental}[2]{\iftoggle{SUPPLEMENTAL}{#1}{#2}}
\newtoggle{BLINDED}\toggletrue{BLINDED}
\togglefalse{BLINDED} % <-------- COMMENT OUT FOR BLINDED VERSION
\newcommand{\Blinded}[2]{\iftoggle{BLINDED}{#1}{#2}}

\usepackage[T1]{fontenc} % for guillemots in Neyman's paper title
\usepackage{graphicx,grffile}
\usepackage{amsmath,amssymb,amsthm}
\usepackage{mathtools,dsfont,centernot}
\usepackage{caption}
\usepackage{booktabs,tabularx,threeparttable,longtable,ragged2e}
\usepackage{siunitx}
\usepackage[shortlabels]{enumitem}
\usepackage[longnamesfirst]{natbib}
\usepackage{hypernat}
\usepackage[nodisplayskipstretch]{setspace}
\usepackage[bottom]{footmisc}
\usepackage{xr} %xr,xr-hyper}

\usepackage[hyphens]{url}
\usepackage{hyperref}

\usepackage{comment}
\usepackage{threeparttable}

\usepackage[final]{listings}
\lstdefinestyle{inlineR}{language=R,frame=none,basicstyle=\ttfamily,keywordstyle=\ttfamily,stringstyle=\ttfamily,keepspaces=true,showspaces=false,showstringspaces=false,breaklines=true,upquote=true,print,columns=fullflexible}
\newcommand{\code}[1]{\lstinline@#1@}

\usepackage[margin=1in,letterpaper]{geometry}
\usepackage[capitalize,noabbrev]{cleveref}

\usepackage{makecell,caption}

  \renewenvironment{thebibliography}[1]%
  {\begin{oldthebibliography}{#1}\setlength{\parskip}{0ex}\setlength{\itemsep}{0ex}}%
  {\end{oldthebibliography}}
\appto\TPTnoteSettings{\linespread{1}\footnotesize}

\DeclareGraphicsExtensions{.pdf,.png}
\graphicspath{ {./Figures/} }

\newcommand{\citeposs}[1]{\citeauthor{#1}'s (\citeyear{#1})}

\crefname{conjecture}{Conjecture}{Conjectures}
\crefname{section}{Section}{Sections}
\crefname{subsection}{Section}{Sections}
\crefname{subsubsection}{Section}{Sections}
\Crefname{conjecture}{Conjecture}{Conjectures}
\Crefname{section}{Section}{Sections}
\Crefname{subsection}{Section}{Sections}
\Crefname{subsubsection}{Section}{Sections}
\crefname{appendix}{Appendix}{Appendices}
\crefname{subappendix}{Appendix}{Appendices}
\crefname{subsubappendix}{Appendix}{Appendices}
\Crefname{appendix}{Appendix}{Appendices}
\Crefname{subappendix}{Appendix}{Appendices}
\Crefname{subsubappendix}{Appendix}{Appendices}
\crefname{equation}{}{}
\Crefname{equation}{Equation}{Equations}
\crefname{enumi}{}{}
\Crefname{enumi}{}{}
\crefname{assumption}{}{}
\Crefname{assumption}{Assumption}{Assumptions}
\Crefname{method}{Method}{Methods}

\theoremstyle{plain}
\theoremstyle{definition}

\newcommand{\vecf}[1]{\boldsymbol{\mathbf{#1}}} %vector formatting

\newcommand{\iid}{\stackrel{\mathit{iid}}{\sim}}

\DeclareMathOperator{\E}{E} %{\mathbb{E}}
\DeclareMathOperator{\1}{\mathds{1}}
\newcommand{\Ind}[1]{\1\{#1\}}

\newcommand{\UnifDist}{\textrm{Unif}}

\newcommand{\diff}{d} %differential operator: easy to change to upright if needed.
\let\originalleft\left
\let\originalright\right
\renewcommand{\left}{\mathopen{}\mathclose\bgroup\originalleft}
\renewcommand{\right}{\aftergroup\egroup\originalright}

\newcommand{\mockalph}[1]{}  % FOR BIBTEX SORTING
\allowdisplaybreaks[3]

\hypersetup{
  pdfauthor = {Anand and Liu},
  pdfkeywords = {economics, econometrics, statistics},
  pdftitle = {Anand and Liu: qdecom cons dist},
  pdfsubject = {econometrics},
  pdfpagemode = UseNone
}

\title{Distributional Decomposition of Consumption Inequality Change During COVID-19}

\author{\Blinded{[BLINDED]}{%
Utkarsh Anand%
\thanks{School of Economic Sciences, Washington State University, Pullman WA, USA, \texttt{utkarsh.anand@wsu.edu}}
\and 
Xin Liu%
\thanks{School of Economic Sciences, Washington State University, Pullman WA, USA, \texttt{xin.liu1@wsu.edu}
}
}}

\date{\today}

\begin{document}

\maketitle

\begin{abstract}

\doublespacing
We decompose the U.S. consumption inequality distributional changes during the COVID-19 phase. 
Analyzing the Consumption Expenditure Interview Survey data, we decompose observed changes in consumption inequality into components attributable to several individual variables.
Using a distribution regression method, we construct counterfactual distributions under the scenario in which the consumption structure or any specific variable would have remained the same between the two years before and after the onset of the COVID-19 pandemic.
We find that changes in the conditional distribution of consumption explain most of the observed decline in consumption inequality among male-headed households between 2018 and 2022.
The rise in asset holdings has significantly increased the consumption inequality in all measures. 
Moreover, the changes in a set of household characteristics have significantly reduced the consumption inequality. 
Our analyses focus on the well-measured consumption components that are robust to the measurement errors in consumption data.

\textit{JEL classification}: 
% https://www.aeaweb.org/jel/guide/jel.php
E21, D63, J11

\end{abstract}

% \pagebreak
\newcommand{\paperspacing}{\onehalfspacing}
\renewcommand{\paperspacing}{\doublespacing}

\paperspacing

\Supplemental{\newpage}{}

\newpage

\section{Introduction}
\label{sec:intro}

Consumption has served as an important indicator of individual economic well-being.
There is a large body of literature and much debate on consumption inequality trends in the U.S. over the past several decades. 
Several studies document a sharp increase in consumption inequality in the 1980s \citep{CutlerEtAl1991,JohnsonShipp1997}, while others report a modest increase during this period \citep{Slesnick1994,Slesnick2001book}. 
In addition, \Citet{FisherJohnsonSmeeding2013} and \citet{MeyerSullivan2013} find evidence of a decline in consumption inequality after 2006.

The COVID-19 pandemic has disrupted the world and the U.S. economy, yielding disparate impacts across socioeconomic groups on outcomes such as employment, educational access, and income. 
The U.S. pandemic-related policy responses also induce heterogeneous effects on the labor market, credit access, borrowing, and financial insecurity. 
In this study, we focus on changes in consumption inequality before and after the COVID-19 pandemic.

In this paper, we add to the consumption inequality literature in three ways. 
First, to the best of our knowledge, this paper is the first in the consumption inequality literature to decompose the observed distributional consumption inequality changes into isolated effects attributable to specific economic factors 
by adopting the counterfactual distribution regression method proposed in \cite{ChernozhukovEtAl2013DR}.
The traditional distributional decomposition method \citep{Melly2005} implemented in most recent consumption inequality studies, such as \cite{MeyerSullivan2023}, decomposes changes in inequality into only three components: characteristics, coefficients, and residuals. 
That is, for example, their ``coefficient effect'' is the aggregate effect of changes in the coefficients of all explanatory variables, which fails to capture the effect of a specific economic factor of interest to researchers. 
Similarly, their ``characteristics effect'' is the aggregate effect of changes in all explanatory variables.
Alternatively, the famous Blinder-Oaxaca decomposition could decompose the observed change due to each explanatory variables, but it limits to the mean of outcome $\E(Y_1)-\E(Y_0)$ and cannot provide the \textit{distributional} decomposition of outcome, for example, $Q_\tau(Y_1)-Q_\tau(Y_0)$, where $Y_1$ and $Y_0$ are two distributions that we want to compare and $\tau \in (0,1)$ is a fixed quantile level.
In this paper, we provide new evidence on how each economic variable contributes to the distributional changes in consumption inequality.
Like the Blinder-Oaxaca decomposition, our distributional decomposition results are statistical rather than causal unless additional assumptions are made. See \citet[Section 2.3]{ChernozhukovEtAl2013DR} for details of conditions under which the decomposition has a causal interpretation.

Specifically, we decompose the observed distributional changes in consumption inequality into components explained by changes in asset holdings, education attainment, family type, work status, other demographic characteristics, and consumption structure (i.e., the conditional distribution of log consumption given all of the explanatory variables of interest).
We find that the changes in the consumption structure explain most of the observed decline in inequality during our sample period.
That is, we compare the counterfactual log consumption distributions holding the household characteristics in the 2018 population distribution, while only letting the conditional distribution of log consumption given explanatory variables change from 2018 to 2022. 
Such a counterfactual change in the log consumption distribution is of a similar magnitude to the observed total change in the log consumption distribution. 
The rise in asset holdings has significantly increased the consumption inequality. Specifically, it helps the middle or poor households less than the upper half of the population.
The increase in higher education attainments has slightly increased the inequality in the lower half of the distribution.
A decrease in the share of households of married couples with no kids has slightly increased the inequality across the distribution. 
Finally, the effects of remaining household characteristics changes altogether have significantly reduced consumption inequality and thus act as an opposing force to assets.

Second, we contribute new evidence on changes in consumption inequality during the COVID-19 pandemic.
Unlike \cite{MeyerMurphySullivan2022} who focus on a very short-run (0--1 year) time frame, 
(i.e., they treat 2020:Q1--2020:Q4 as the sampled time periods to compare with the pre-COVID periods 2019:Q1--2019:Q4, and thus they observe a dramatic decline in consumption), we focus on a 3--5 year horizon at which point the economy has recovered mostly 
and we contribute new findings on the mid-run impacts of the pandemic,
filling a gap in the literature.

The COVID-19 pandemic has both short-term and long-term effects on consumption through different channels. 
The short-term effect of COVID-19 on consumption could be more likely due to lockdowns and physical constraints that limit regular in-person consumption at restaurants, gas stations, and grocery stores, as well as financial constraints and liquidity issues that prompt households to build precautionary savings.

Over a longer time span, the effect of COVID-19 may set in through various mechanisms and impact consumption heterogeneously across households and across years.
First, the shock to consumption and the subsequent rebound of consumption would differ across different stages during and after COVID.  
In 2022, there may still be a lot of catch-up consumption (for example, traveling) that was delayed during 2020-2021.
Second, consumption is closely related to saving, and the saving behaviors also change dramatically during and after the COVID-19 pandemic.
When COVID hit in 2020, households with or without savings and other borrowing resources to smooth out the shock may adjust their consumption heterogeneously. 
Since the pandemic, more households have begun to build precautionary savings, which substitute for some consumption in the short run but may have different long-run effects.
Third, the asset distribution has changed a lot within a few years, and in a way that substantially changed the consumption conditional on assets.
Finally, the pandemic has brought lifestyle changes that also take years to affect one's consumption.
For example, working from home may permanently boost the household's demand for housing upgrades. 
Many households have experienced geographic relocation, which has affected their commutes, gas costs, and regional consumption patterns. 
And people who increase alcohol consumption or get depression caused by isolation during the pandemic may have long-term health consequences and changes in their consumption patterns. 
These factors could affect consumption differently. 
Our method, in principle, can isolate the effect of each factor when such variables are available in the data. 
We focus on assets and a few demographic variables, as well as the conditional distribution of log consumption given these explanatory variables in our analysis.

Specifically, we consider a pre-COVID-19 time period (2018:Q4) and a post-COVID-19 time period (2022:Q4). 
We compare the household-level log consumption distributions and changes in consumption inequality between these two years. 
We find that the consumption inequality has decreased modestly in the upper half ($90{:}50$ ratio) of the consumption distribution, while it has decreased slightly in the lower half ($50{:}10$ ratio).

Third, as a minor contribution, we impute the well-measured components of consumption and leverage them for the computation of inequality statistics, following \citeposs{MeyerSullivan2023} recommendations. 
It is well known that the observed total consumption includes poorly measured components that suffer from measurement errors and under-reporting issues, and that the reporting quality is declining over time. 
Such reporting issues make the inequality statistics computed using the observed total consumption biased and unreliable.
\cite{MeyerSullivan2023} advocate the well-measured consumption components instead.
They show that inequality statistics calculated from well-measured components can closely approximate the consumption inequality of the true latent total consumption under certain assumptions. 
Among their assumptions for leveraging the well-measured consumption, two are testable.
We examine those testable assumptions for our pre- and post- COVID-19 time period samples, following \citeposs{MeyerSullivan2023} method.

There is a rapid growth of literature on the effect of the COVID-19 pandemic on the socioeconomic outcomes, including employment \citep{HershbeinHolzer2021}, income \citep{Kollar2023}, and education \citep{LeeEtAl2021}, among others. 
Employment and income are closely related and are immediately affected by COVID-19, which is then transmitted to consumption. 
Workers in high-contact, low-paid jobs, such as those in restaurants and hotels, are often hit by COVID immediately. 
Occupations that are easier to transition to work-from-home during COVID-19 tend to require higher levels of computer skills and are usually higher-paid jobs.
Higher-income households tend to have more savings and assets that help them maintain inelastic consumption, although their elastic luxury consumption may be substantially reduced.
The impact of COVID-19 on education, then on one's occupation and income, and then on consumption may take longer to reveal. 
The school closures during COVID-19 may disproportionately affect K-12 students, especially those from lower-income families, who may have difficulty accessing online education due to a lack of a home computer and/or internet access. 
These effects on education may take longer to translate into consumption changes, as it may take a few years for students to finish their college education and enter the labor force, and this may affect their occupation and income.
See \cite{MeyerEtAl2025} for a review.

The remainder of the paper is organized as follows. 
Data and methods are reported in \cref{sec:data,sec:method}.
\cref{sec:emp} presents the results and \cref{sec:conclusion} concludes.

\section{Data}
\label{sec:data}

\subsection{Data sources and variables}

We use household-level data from the Consumer Expenditure Interview Survey released by the U.S. Bureau of Labor Statistics.
Our sample consists of two repeated cross-sections: 2018:Q4 and 2022:Q4. 
As noted in \cite{AttanasioPistaferri2017}, the Consumer Expenditure Survey is the only publicly available dataset with detailed information on household expenditure in the U.S. 

We focus on the well-measured consumption components for our inequality study as suggested by \cite{MeyerSullivan2023}.
It includes the categories of expenditures on food at home, utilities, gasoline and oil, housing, and vehicles. 
We impute the quarterly rental equivalent values as housing expenditure, and the quarterly service flow as expenditure on vehicles using the imputation method proposed in \cite{MeyerSullivan2023}.
The imputation details are presented in \cref{sec:impute}.
\cite{MeyerSullivan2023} and their prior work \cite{BeeEtAl2015} report that the ratio of the well-measured consumption categories relative to the national accounts has been high (close to one) and stable over time.
We briefly review \citeposs{MeyerSullivan2023} consumption model in \cref{secsub:ConsModel} and their assumptions in \cref{secsub:ConsAssp}.
Under these assumptions, the consumption inequality statistics computed using the well-measured components can reliably approximate the true inequality statistics from the latent total consumption.
We examine whether these assumptions hold for our data in \cref{subsec:ConsAsspVerify}.

We decompose observed changes in consumption inequality into contributions from various economic variables. 
Specifically, we consider household asset holdings, educational attainment, employment status, and other household characteristics. 
The household asset holdings include financial assets (cash, stocks, bonds), real estate, pension funds, and vehicle values, net of mortgages and other debt. 
Asset holdings reflect a household’s capacity to smooth consumption over time, especially in the face of income disruptions. 
Higher education leads to higher skills, more employment opportunities, and greater earning potential. 
It is relatively easy for people with higher education to adapt to remote work during COVID. 
Thus, they are less impacted by COVID shocks than people with lower levels of education and/or in lower-paid occupations. 
Employment status directly reflects whether a household has a stable income to support monthly consumption. 
Other household characteristics include the age of the head of household, family types based on marital status and child status, race, and urban or rural residence. 

Like \cite{AguiarBils2015} who studied consumption inequality, we focus on log consumption and further decompose its change into factors such as log assets. 
This log model forces us to 
remove observations with missing, zero, or negative values in consumptions and assets, which results in 3898 male-headed households in total, (2023 observations in year 2018:Q4 and 1875 in year 2022:Q4).\footnote{
There are 4951 observations in our two-year pooled male-headed raw sample. 
Among them, 558 observations have missing values of consumption,  
201 observations have zero values in assets, and
597 observations have negative asset values, with some observations overlapping across these categories.} 
This implementation is common in the study of consumption and earning, for example, \citet[p.\ 712]{ArellanoEtAl2017} who study the consumption and earning dynamics drop observations in the same way.\footnote{They note ``we drop all observations for which data on earnings, consumption, or assets, either in levels or log-residuals, are missing.''}
Admittedly, such a log model may be less appropriate when a large fraction of observations report zero consumption or negative asset values. 
A careful consideration of the missing data and sample selection issue is left for future work.

We focus on male-headed households in our main analysis and present the results for female-headed households in \cref{sec:female}.
We separate the analysis for male- and female-headed households, 
because female-headed households are often headed by single mothers or widows and tend to face social disadvantages, such as lower wages and higher poverty rates. 
Female-headed households may have different consumption patterns from male-headed households.
Our sample further confirms these gender-based economic disparities: male-headed and female-headed households differ across many characteristics, including consumption levels, asset holdings, and education and family characteristics. 
For example, real assets have increased by only 13.2\% for female-headed households, but by more than twice as much, 32.9\%, for male-headed households. 
The single mother with the oldest kid under 18 accounts for 8 percent of the female-headed households, whereas the single father with the oldest kid under 18 accounts for only 3 percent of the male-headed households. 
In addition, the female-headed households exhibit a different pattern of change than the male-headed households between the two years. 
Pooling male-headed and female-headed households together would make the gender-characteristics specific effects all wrapped into a single gender-difference effect.

\subsection{Summary statistics}

\begin{table}[htbp]
\centering
\sisetup{round-mode=places,round-precision=2,detect-all}
\begin{threeparttable}
\caption{Descriptives for male-headed household consumptions and assets}
\label{tab:stat:male:1}
\begin{tabular}{l
S[table-format=2.2]
S[table-format=3.2]
S[table-format=2.2]
S[table-format=4.2]
c}
\toprule
\multicolumn{6}{c}{Year 2018} \\
\midrule
{Variables} & {Mean} & {SD} & {Min} & {Max} & {N}  \\
\midrule
Nominal consumption (in \$1000) & 7.861676 & 4.290335 & 0.840844 & 42.610582 & 2023 \\
Real consumption (in \$1000)    & 3.130807 & 1.708568 & 0.334855 & 16.969094 & 2023 \\
Log real consumption            & 7.924843 & 0.495255 & 5.813697 & 9.739149 & 2023 \\
Nominal asset holdings (in \$1000) & 39.236041 & 128.211091 & 0.000616 & 4011.397000 & 2023 \\
Real asset holdings (in \$1000)    & 15.625228 & 51.058350 & 0.000245 & 1597.485125 & 2023 \\
Log real asset holdings            & 8.859214 & 1.131576 & -1.405912 & 14.283941 & 2023\\
\addlinespace
\midrule
\multicolumn{6}{c}{Year 2022} \\
\midrule
{Variables} & {Mean} & {SD} & {Min} & {Max} & {N} \\
\midrule
Nominal consumption (in \$1000) & 10.003866 & 5.103485 & 1.430774 & 43.381695 & 1875 \\
Real consumption (in \$1000)    & 3.418314  & 1.743857 & 0.488894  & 14.823493 & 1875 \\
Log real consumption            & 8.021926  & 0.481749 & 6.192146  & 9.603969 & 1875  \\
Nominal asset holdings (in \$1000) & 60.790864 & 221.070199 & 0.014000 & 3939.945500 & 1875 \\
Real asset holdings (in \$1000)    & 20.772194 & 75.539526  & 0.004784 & 1346.276500 & 1875 \\
Log real asset holdings            & 8.975021  & 1.114652 & 1.565233 & 14.112853 & 1875\\
\bottomrule
\end{tabular}
\begin{tablenotes}[para,flushleft]
\footnotesize{}
The consumptions here are only the well-measured components.
Base year 1982--1984=100.
\end{tablenotes}
\end{threeparttable}
\end{table}

We incorporate the survey weights for summary statistics in \cref{tab:stat:male:1,tab:stat:male:2}.
Following \citeposs{MeyerSullivan2023}, we adjust for the differences in family size to reflect the consumption differences between adults and children and reflect the diminishing marginal cost of an additional adult.

\Cref{tab:stat:male:1} presents the consumption and asset variables for the male-headed households. 
Real consumption and real asset holdings are adjusted for inflation using a base period of 1982--1984$=100$, based on the All Items CPI. (Other price indices are presented in \cref{tab:inflation_indices} with a discussion of which results will be affected and which not if using a different price index.)
The average quarterly household nominal consumption has increased by 27.2\% from \$7{,}862 in 2018 to \$10{,}004 in 2022. 
The real consumptions have increased by 9.2\% from \$3{,}131 in 2018 to \$3{,}418 in 2022. 
Both the nominal and real consumption growth match the national accounts.
The standard deviation (SD) of real consumption has increased modestly by about 2.0\% from \$1{,}709 to \$1{,}744, while the SD of the log real consumption has decreased modestly by about 1.3\% from 0.495 to 0.482. 
This indicates that the real consumption is right-skewed and that its SD is heavily influenced by extreme values.
We focus on the log real consumption, which helps reduce skewness.

The average quarterly household nominal asset holdings have increased by 54.9\% from \$39{,}236 in 2018 to \$60{,}791 in 2022. 
And the real assets have increased by 32.9\% from \$15{,}625 in 2018 to \$20{,}772 in 2022. 
The significant increase in assets is not surprising, as both the US stock market and real estate market have experienced skyrocketing asset values during this period.
The SD of real assets has increased substantially by 47.9\% ($=75.54/51.06-1$)  %from \$51{,}058 to \$75{,}540,
while the SD of log real assets has decreased modestly by 1.5\% ($=1.115/1.132-1$). 
This indicates that the real asset is right-skewed; thus, we focus on the log of the real asset to mitigate the skewness.

\begin{figure}[ht]%[htbp]
\centering
\scriptsize
\includegraphics[width=\textwidth]{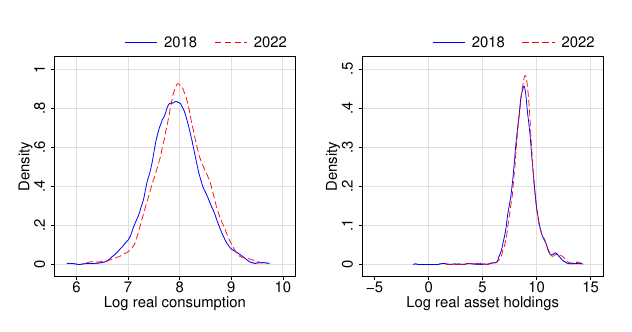}
\caption{Density curves for log real consumption (left panel) and log real assets (right panel) for male-headed households}
\label{fig:density:male}
\end{figure}

\Cref{fig:density:male} presents the kernel density curves of log real consumption and log real assets in 2018 and 2022. 
The density curve for the log real consumption has shifted modestly to the right, and the 2018 curve looks slightly more spread out than the 2022 curve. This is consistent with the increase (from 7.925 to 8.022) in the mean and a decrease (from 0.495 to 0.481) in the SD of log real consumption in \cref{tab:stat:male:1}.
The density curve for log real assets has also shifted modestly to the right, consistent with the increase in its mean.

\begin{table}[ht]
\centering
\def\sym#1{\ifmmode^{#1}\else\(^{#1}\)\fi}
\sisetup{round-precision=2,round-mode=places,detect-all}
\begin{threeparttable}
\caption{Summary statistics for individual characteristics of the male-headed household heads}
\label{tab:stat:male:2}
\begin{tabular}{l
S[table-format=3.2,round-precision=2]
S[table-format=3.2,round-precision=2]
c
S[table-format=3.2,round-precision=2]
S[table-format=3.2,round-precision=2]
c
S[table-format=3.2,round-precision=2]
S[table-format=3.2,round-precision=2]}
\toprule
& \multicolumn{2}{c}{2018} && \multicolumn{2}{c}{2022} && \multicolumn{2}{c}{Difference} \\
\cmidrule{2-3} \cmidrule{5-6} \cmidrule{8-9}
{Variables} & {Mean} & {SD} && {Mean} & {SD} && {Mean} & {SE} \\
\midrule
Not at work &  0.200244& 0.400282  && 0.199853&  0.399997 && -0.000391 & 0.014396 \\
Age & 49.210485& 15.898451 && 49.741681& 16.033030 &&  0.531196 & 0.603262 \\ [2pt]
Married, no kids and no others & 0.255851&  0.436446 && 0.221884& 0.415624 &&  -0.033967\sym{**} & 0.014452 \\
Married, oldest child age $<$6 & 0.080492& 0.272121 && 0.057448& 0.232758 &&  -0.023045\sym{**} &  0.010571 \\
Married, oldest child age 6--17  & 0.210283& 0.407610  && 0.216975& 0.412295 &&  0.006692 & 0.018960 \\
Married, oldest child age $>$17  & 0.117594& 0.322207 && 0.128033& 0.334216 && 0.010439 & 0.015263 \\
Married, others  &  0.107666& 0.310034 && 0.095729& 0.294297 && -0.011937 &  0.017478 \\
Single father, oldest child age $<$18 & 0.025495& 0.157663 && 0.020463& 0.141616 &&  -0.005032 & 0.005734 \\
Single male & 0.095623& 0.294147 && 0.112368& 0.315902 &&  0.016744\sym{**} &  0.007442 \\
None of the above family types & 0.106995& 0.309184 && 0.147101& 0.354302 && 0.040106\sym{***} & 0.013461 \\ [2pt]
% No schooling or elementary & 0.039169& 0.168323 && 0.035491& 0.185067 && 0.006322 &  0.008673 \\
High school graduate & 0.288233& 0.453052 && 0.239992& 0.427192 &&   -0.048240\sym{**} & 0.018843 \\
Some college & 0.257590& 0.437415 && 0.267748& 0.442904 && 0.010159 & 0.017569 \\
Bachelor's degree or above & 0.414008& 0.494466 && 0.456768& 0.498260  &&  0.042760\sym{**} & 0.0159 \\
White & 0.671405& 0.469819 && 0.640031& 0.480119 && -0.031374 & 0.021065 \\
Black & 0.106174& 0.308137 && 0.100421& 0.300641 &&  -0.005753 & 0.016247 \\
Other races & 0.222421& 0.415975 && 0.259548& 0.438503 &&   0.037127\sym{**} &  0.018385 \\
Urban area & 0.939826& 0.237867 && 0.933892& 0.248536 && -0.005934 &  0.011134 \\
\midrule
Number of observations & \multicolumn{2}{c}{2023} && \multicolumn{2}{c}{1875} && \multicolumn{2}{c}{} \\
\bottomrule
\end{tabular}
\begin{tablenotes}[para,flushleft]
\footnotesize{}
\item  
\textsuperscript{***}~p<0.01, \textsuperscript{**}~p<0.05, \textsuperscript{*}~p<0.1.
The ``married, others'' represents married couples living with individuals beyond their children. 
\end{tablenotes}
\end{threeparttable}
\end{table}

\Cref{tab:stat:male:2} provides summary statistics for individual-level characteristics for the head of households. 
We focus on the male-headed households for the reasons mentioned above.
All variables, except age, are dummy variables.
The variable ``not at work'' takes the value 1 if the individual has never worked (full or part-time) in the past 12 months, 
which includes those who are unemployed, retired, and/or out of the labor force.
Only 2837 out of the total 3898 male-headed observations have occupation information.
We could not differentiate between remote work and on-site work in our sample.
The proportion of not-at-work household heads has remained roughly the same between the two years.
The age ranges from 18 to 88. 
In the sample, 0.8\% are aged 18--22, and 37.7\% are aged 59--88.
The proportion of married couples without kids has decreased by 3 percentage points, and 
the share of families of a married couple with a child less than 6 years old has shrunk by 2 percentage points.
The proportion of the population with only a high school degree and not even some college has decreased by 5 percentage points, while the proportion with a bachelor’s degree or above has increased by 4 percentage points.

\section{Method}
\label{sec:method}

This section presents \citeposs{ChernozhukovEtAl2013DR} distribution regression and decomposition methods adopted for our analysis.
We construct counterfactual distributions, present the decomposition, compute the inequality statistics, and discuss how this method differs from and potentially improves upon the traditional decomposition approaches to analyzing consumption inequality.

\subsection{Counterfactual distribution and decomposition}
\label{subsec:method}

Let 18 denote the population in the year 2018 (before the COVID-19 period) and let 22 denote the population in the year 2022 (post-COVID-19 period). 
Let $F_{Y_{18} \mid X_{18}}(y\mid \vecf{x})$ and $F_{Y_{22} \mid X_{22}}(y\mid \vecf{x})$ denote the conditional distribution function of consumption given the characteristics $\vecf{x}$ in year 2018 and 2022, respectively. 
Let $F_{Y_{18 , 18}}$ and $F_{Y_{22 , 22}}$ denote the observed unconditional distributions of log real consumption in these two years.
Let $F_{Y_{18 , 22}}$ denote the counterfactual unconditional log real consumption distribution in 2022 if households in 2022 were facing the year 2018's consumption structure function. 
That is, 
\begin{equation}
\label{eqn:simple:ctfn}
    F_{Y_{18 , 22}} = \int_{\mathcal{X}} F_{Y_{18} \mid X_{18}}(y\mid \vecf{x}) \ d F_{X_{22}}(\vecf{x}) ,
\end{equation}
where $\mathcal{X}$ is the support of household characteristics, which is assumed to be the same across the two years $\mathcal{X}_{18}=\mathcal{X}_{22}$;
$F_{X_{22}}$ is the covariates distribution in year 2022;
and the conditional distribution $F_{Y_{18} \mid X_{18}}(\cdot \mid \vecf{x})$ can be estimated via a collection of binary dependent variable regressions using the 2018 sample.\footnote{We use logistic regression for this intermediate step in our empirical results.
\cite{ChernozhukovEtAl2013DR} provides a robustness check to different binary dependent variable regressions: logit, probit, linear probability model, Cauchit, and complementary log-log in their empirical example.
They find that the differences in the final decomposition results for adopting different binary outcome regression methods are ``so modest and almost indistinguishable.'' 
}

The observed log consumption distribution differences between these two years can be decomposed as
\begin{equation}
\label{eqn:simple:decom}
    \underbrace{F_{Y_{22 , 22}} - F_{Y_{18 , 18}} }_{\textrm{observed} }
    =  \underbrace{F_{Y_{22 , 22}}-F_{Y_{18 , 22}} }_{\textrm{price effect} } + \underbrace{F_{Y_{18 , 22}}-F_{Y_{18 , 18}} }_{\textrm{composition effect}} ,
\end{equation}
where the first term on the right-hand side measures a ``price effect'' which is the hypothetical change of outcome distribution due to the differences in consumption structures (i.e., the different conditional distributions of consumption given the covariates in the two years respectively) while keeping the characteristics in the same year 2022.
Note that for each year $t \in \{18, 22\}$, 
the ``consumption structure'' refers to not merely the conditional mean of log consumption, but rather a collection of the conditional distributions of log consumption at every threshold value in the domain of log consumption $F_{Y_{t}|X_{t}}(y \mid  X_{t}) = \E[\Ind{Y_{t} \leq y} \mid \vecf{X}_{t}]= \vecf{X}_{t}'\vecf{\gamma}_t(y)$
at every $y \in \mathcal{Y}$, 
and the conditional distribution coefficients $\vecf{\gamma}_t(y)$ are allowed to vary across the threshold values $y$. 
The second term on the right hand side of \cref{eqn:simple:decom} measures the ``composition effect'' which is the hypothetical change of log consumption distribution due to the population characteristics changes from $X_{18}$ to $X_{22}$ while keeping the conditional distributions of consumption $\{F_{Y_{18}|X_{18}}(y \mid X_{18}), \forall y \in \mathcal{Y}\}$ unchanged in 2018.

Note that the decomposition results may differ when replacing $F_{Y_{18 , 22}}$ in \cref{eqn:simple:decom} with $F_{Y_{22 , 18}}$.
The decomposition with a specific counterfactual distribution is associated with a specific economic meaning. 
For example, using the counterfactual distribution $F_{Y_{22 , 18}}$ in the decomposition, 
the $F_{Y_{22 , 18}}-F_{Y_{18 , 18}}$ measures the ``price effect'' of changing from 2018's consumption structure to 2022's structure faced by the 2018 population.
(And $F_{Y_{22 , 22}}-F_{Y_{22 , 18}}$ measures the ``composition effect'' of changing the population characteristics from $X_{18}$ to $X_{22}$ while keeping the conditional distributions of log consumption in 2022.) 
Compared to \cref{eqn:simple:decom}, the $F_{Y_{22 , 22}}-F_{Y_{18 , 22}}$ measures the ``price effect'' of changing from 2018's consumption structure to 2022's structure faced by the 2022 population.
These two ``price effects'' should be qualitatively similar, as they both measure the counterfactual changes from the 2018 consumption structure to that of 2022.
They differ, however, both numerically and economically, in that the counterfactual change is applied to the 2018 population versus the 2022 population.

Below, we illustrate how to decompose the observed differences in log consumption distribution across the two years into multiple explanatory variables, including a factor $X_1$, a factor $X_2$, a vector of all other characteristics $\vecf{X}_3$, and the consumption structure. 
(It can be easily extended to a decomposition into every single explanatory variable of $X_1, \ldots, X_k$ as well as the consumption structure.)

Let $F_{Y_{(t,s,r,v)}}$ denotes the counterfactual (unconditional) consumption distribution, 
when the conditional consumption structure is in year $t$, the factor $X_1$ in year $s$, the factor $X_2$ in year $r$, and all the other characteristics in year $v$. The $\{t,s,r,v\}\in \{2018, 2022\}$.
The observed unconditional consumption differences between the two years can be decomposed as four counterfactual distributional effects (DE),
\begin{align}
\label{eqn:DE:decom}
    F_{Y_{22, 22, 22, 22}} -F_{Y_{18, 18, 18, 18}}
    & = \underbrace{F_{Y_{22, 22, 22, 22}} - F_{Y_{22, 18, 22, 22}}}_{\textrm{DE due to factor } X_1}
        + \underbrace{F_{Y_{22, 18, 22, 22}} - F_{Y_{22, 18, 18, 22}} }_{\textrm{DE due to factor } X_2} \notag \\
    & \quad + \underbrace{F_{Y_{22, 18, 18, 22}} - F_{Y_{22, 18, 18, 18}} }_{\textrm{DE due to other characteristics } \vecf{X}_3}
        + \underbrace{F_{Y_{22, 18, 18, 18}} - F_{Y_{18, 18, 18, 18}} }_{\textrm{DE due to consumption structure}}.
\end{align}

Similarly, consider the quantile function to be the left inverse of the distribution function. 
We can decompose the observed quantile function differences into four counterfactual quantile effects (QE),
\begin{align}
\label{eqn:QE:decom}
    & Q_{Y_{22, 22, 22, 22}}(\tau) -Q_{Y_{18, 18, 18, 18}}(\tau) \notag \\
    & = \underbrace{Q_{Y_{22, 22, 22, 22}}(\tau) - Q_{Y_{22, 18, 22, 22}}(\tau)}_{\textrm{QE due to factor } X_1} 
        + \underbrace{Q_{Y_{22, 18, 22, 22}}(\tau) - Q_{Y_{22, 18, 18, 22}}(\tau) }_{\textrm{QE due to factor } X_2} \notag \\
    & \quad + \underbrace{Q_{Y_{22, 18, 18, 22}}(\tau) - Q_{Y_{22, 18, 18, 18}}(\tau) }_{\textrm{QE due to other characteristics } \vecf{X}_3}
        + \underbrace{Q_{Y_{22, 18, 18, 18}}(\tau) - Q_{Y_{18, 18, 18, 18}}(\tau) }_{\textrm{QE due to consumption structure}}, 
\end{align}
where $\tau \in (0,1)$ denotes a fixed quantile level.

The counterfactual distributions are constructed by integrating the observed conditional consumption distribution with respect to the hypothetical distribution of characteristics. 
It can be estimated using the plug-in principle.
Specifically, the three counterfactual distributions in \cref{eqn:DE:decom} are constructed as follows,
\begin{align*}
    F_{Y_{22,18,22,22}}(y) 
    & = \iint F_{Y_{22} \mid X_{1}^{22}, X_{2}^{22}, \vecf{X}_{3}^{22}} (y \mid x_1, x_2, \vecf{x}_{3}) \, dF_{X_{1}^{18} \mid X_{2}^{18}, \vecf{X}_{3}^{18}}(x_1 \mid x_2, \vecf{x}_{3}) \, dF_{X_{2}^{22}, \vecf{X}_{3}^{22}}(x_2, \vecf{x}_{3}) ; \\
    F_{Y_{22,18,18,22}}(y) 
    &= \iiint F_{Y_{22} \mid X_{1}^{22}, X_{2}^{22}, \vecf{X}_{3}^{22}} (y \mid x_1, x_2, \vecf{x}_{3}) \, dF_{X_{1}^{18} \mid X_{2}^{18}, \vecf{X}_{3}^{18}}(x_1 \mid x_2, \vecf{x}_{3}) \notag \\
    &\quad dF_{X_{2}^{18} \mid \vecf{X}_{3}^{18}}(x_2 \mid \vecf{x}_{3}) \, dF_{\vecf{X}_{3}^{22}}(\vecf{x}_{3}) ;  \\
    F_{Y_{22,18,18,18}}(y) 
    & = \int F_{Y_{22} \mid X_{1}^{22}, X_{2}^{22}, \vecf{X}_{3}^{22}}(y \mid x_1, x_2, \vecf{x}_{3}) \, dF_{X_{1}^{18}, X_{2}^{18}, \vecf{X}_{3}^{18}}(x_1, x_2, \vecf{x}_{3}) .
\end{align*}

As mentioned in \cref{sec:data}, we decompose the observed log consumption distribution changes due to changes in the household asset holdings, education attainments, employment status, and other household characteristics. 
We focus on the effects of changes in household asset holdings, education attainment, and the remaining explanatory variables together in \cref{sec:emp}. 
We further separate the effects of changes in one specific family type and the work status from the remaining covariates in \cref{sec:emp:extra}.

\subsection{Inequality statistics}

We consider seven inequality measures: the standard deviation and the Gini coefficient of log consumption ($Y \equiv \ln(C)$), and the five (log-transformed) inequality ratios of consumption, including the $90{:}10$ ratio, $90{:}50$ ratio, $50{:}10$ ratio, $75{:}25$ ratio, and the $95{:}5$ ratio.

The inequality ratio is measured as the log of the ratio, or equivalently, as the difference between the two quantiles of log consumption.
In general, the $b:a$ interquantile range of the log consumption distribution is the log of the $b:a$ ratio of consumption.
For example, 
$Q_{0.9}(\ln(C)) - Q_{0.1}(\ln(C))
= \ln(Q_{0.9}(C)) - \ln(Q_{0.1}(C))
= \ln(Q_{0.9}(C)/Q_{0.1}(C))
= \ln(\textrm{90:10 ratio of } C)$.
Given two log consumption distributions $\ln(C_1)$ and $\ln(C_0)$,
we use the differences $[Q_{0.9}(\ln(C_1)) - Q_{0.1}(\ln(C_1))]-[Q_{0.9}(\ln(C_0)) - Q_{0.1}(\ln(C_0))]=\ln(\textrm{90:10 ratio of } C_1)-\ln(\textrm{90:10 ratio of } C_0)$ to measure the percentage change of $b:a$ consumption ratio.
Since $\ln(W_1)-\ln(W_0)\approx (W_1-W_0)/W_0$ for any random variable $W$ when the change $W_1-W_0$ is small, the log difference multiplied by 100 approximately measures the percentage change of the original variable. 
The same implementations also appear in \cite{MeyerSullivan2023} and \cite{ChernozhukovEtAl2013DR} for studying the change in the inequality ratio with a log outcome. See \citet[Appendix D]{MeyerSullivan2023} for details.

The Gini coefficient is defined as the area between the $45$-degree line of perfect equality and the Lorenz curve. 
For the counterfactual log consumption distribution $Y_{(t,s,r,v)}$, its Gini coefficient is computed as
\begin{equation*}
\textrm{Gini}_{(t,s,r,v)} = 1 - 2 \int L(y, F_{Y_{(t,s,r,v)}} ) \, dF_{Y_{(t,s,r,v)}}(y) ,
\end{equation*}
where the Lorenz curve is defined as
\begin{equation*}
L(y, F_{Y_{(t,s,r,v)}}) = \int \Ind{\tilde{y} \leq y} \cdot \tilde{y} \, dF_{Y_{(t,s,r,v)}}(\tilde{y}) 
\bigg/
\int \tilde{y} \, dF_{Y_{(t,s,r,v)}}(\tilde{y}) .
\end{equation*}

\subsection{Compare with traditional decomposition method}

The decomposition method in \cref{subsec:method} is fundamentally different from the traditional methods.
Traditional methods usually decompose the changes between two observed distributions into ``coefficients'' (the pooled effects of changes in all coefficients), ``characteristics'' (the pooled effects of changes in all characteristic variables), and ``residuals'' (the remaining unexplained part).
Instead, we decompose the observed total log consumption changes into components attributable to the consumption structure change and the change in each explanatory variable.
We construct the counterfactual distribution by keeping all other covariates and the consumption structure the same across the two years, and letting only the variable of interest change. 
We compare the counterfactual distributions induced by this single-variable change.

For comparison, we review in \cref{subsec:MellyModel} the traditional decomposition method \citep{Melly2005} and discuss its differences and advantages, as well as those of the recently developed method \citep{ChernozhukovEtAl2013DR} that we adopt for our analysis.
In short, the traditional method relies on many assumptions that are less likely to hold in practice, resulting in biased estimated effects. 
Our adopted method allows for a flexible nonlinear conditional quantile function of consumption; thus, our estimated effects are arguably more reliable, although this comes at the expense of extensive computations to analytically trace out the true conditional quantile function of consumption and build on that to construct the unconditional counterfactual consumption distributions.

\section{Results}
\label{sec:emp}

\begin{table}[htbp]
\centering
\caption{\label{tab:emp:male:4factor}Decomposition of consumption inequality changes}
\sisetup{round-precision=1,round-mode=places,detect-all}
\begin{threeparttable}
\begin{tabular}{lccccc}
\toprule
& & \multicolumn{4}{c}{Effects of changes in} \\
\cmidrule{3-6}
Statistics & {\makecell{Total\\Changes}} & {Assets} & {\makecell{College\\Education }} & {\makecell{Remaining\\ Covariates}} & {\makecell{Consumption\\ Structure}} \\
\midrule

SD          
&  \num{-1.544154} (\num{1.499138}) 
& \num{4.375745} (\num{2.055125}) 
& \num{0.298745} (\num{0.222765})
& \num{-4.682189} (\num{3.198376}) 
& \num{-1.536455} (\num{10.028599}) 
\\[0.65em]

% Share estimates: Total=.; Assets=-283.37493; Education=-19.346865; Remaining=303.22036; Structure=99.501429
% Share SEs: Total=.; Assets=437.0776;  Education=24.10748;   Remaining=449.43491; Structure=540.14785

90--10      
&  \num{-4.490374} (\num{5.543086}) 
& \num{11.923610} (\num{4.212275}) 
& \num{1.173945} (\num{0.673783}) 
& \num{-13.304721} (\num{5.282971}) 
& \num{-4.283209} (\num{39.091794}) 
\\[0.65em]

% Share estimates: Total=.; Assets=-265.53714; Education=-26.143597; Remaining=296.29428; Structure=95.386459
% Share SEs: Total=.; Assets=567.48181; Education=10.992001;  Remaining=341.50311; Structure=371.87821

50--10      
&  \num{-0.801027} (\num{4.764475}) 
& \num{7.318913} (\num{3.348183}) 
& \num{1.283847} (\num{0.683176}) 
& \num{-8.036822} (\num{4.759811}) 
& \num{-1.366965} (\num{7.922596}) 
\\[0.65em]

% Share estimates: Total=.; Assets=-913.6908;  Education=-160.27513; Remaining=1003.3144; Structure=170.65148
% Share SEs: Total=.; Assets=406.48232; Education=18.988753;  Remaining=353.70363; Structure=304.45722

90--50      
&  \num{-3.689347} (\num{2.630227}) 
& \num{4.604698} (\num{1.772209}) 
& \num{-0.109902} (\num{0.344584}) 
& \num{-5.267898} (\num{1.866808}) 
& \num{-2.916244} (\num{3.047758}) 
\\[0.65em]

% Share estimates: Total=.; Assets=-124.81066; Education=2.978908;  Remaining=142.78676; Structure=79.044992
% Share SEs: Total=.; Assets=167.94603; Education=10.619623; Remaining=157.40933; Structure=61.660091

75--25      
&  \num{-2.744663} (\num{2.997578}) 
& \num{6.792994} (\num{2.020490}) 
& \num{0.435589} (\num{0.388639}) 
& \num{-7.407987} (\num{2.895216}) 
& \num{-2.565259} (\num{5.013110}) 
\\[0.65em]

% Share estimates: Total=.; Assets=-247.49829; Education=-15.870415; Remaining=269.90518; Structure=93.463524
% Share SEs: Total=.; Assets=379.84419; Education=14.513418;  Remaining=304.62395; Structure=119.4104

95--5       
&  \num{-7.520209} (\num{7.205908}) 
& \num{13.112467} (\num{7.118399}) 
& \num{0.921580} (\num{0.713682}) 
& \num{-12.463363} (\num{12.694313}) 
& \num{-9.090894} (\num{32.054363}) 
\\[0.65em]

% Share estimates: Total=.; Assets=-174.36307; Education=-12.254715; Remaining=165.7316;  Structure=120.88618
% Share SEs: Total=.; Assets=372.50263; Education=7.07796;    Remaining=302.52616; Structure=324.31608

Gini        
&  \num{-0.424456} (\num{0.304382}) 
& \num{0.928373} (\num{0.345731}) 
& \num{0.064096} (\num{0.046938}) 
& \num{-1.030835} (\num{0.571801}) 
& \num{-0.386090} (\num{1.570942}) 
\\

% Share estimates: Total=.; Assets=-218.72074; Education=-15.100769; Remaining=242.86039; Structure=90.961119
% Share SEs: Total=.; Assets=211.07714; Education=16.505974;  Remaining=331.37384; Structure=337.35556

\bottomrule
\end{tabular}
\begin{tablenotes}[para,flushleft]
\footnotesize{}
Male-headed households only.
All values are in percentages. 
Bootstrapped standard errors with 500 repetitions are reported in parentheses, which follows Algorithm 3 in \cite{ChernozhukovEtAl2013DR}.
\end{tablenotes}
\end{threeparttable}
\end{table}

\Cref{tab:emp:male:4factor} reports the changes of the seven inequality statistics between 2018 and 2022 and the decomposition of these changes into four effects, due to the changes in the household asset holdings, education attainments, other household characteristics, and the consumption structure.

For the observed total changes,
the consumption inequality has decreased by 0.8\% for the lower half of the distribution ($50{:}10$ ratio) and decreased modestly by 3.7\% for the upper half of the distribution ($90{:}50$ ratio). 
That is, consumption inequality has shrunk less on the lower half of the distribution than on the upper half.
Together the $90{:}10$ ratio has decreased by 4.5\% ($=0.8\%+3.7\%)$.
When looking at a broader range of the distribution ($95{:}5$ ratio), consumption inequality has dropped by 7.5\%, a larger decline than for the $90{:}10$ ratio.
This shows that the inequality between the two ends of the consumption distribution actually shrinks. 
The inequality for the middle half of the population ($75{:}25$ ratio) has dropped by 2.7\%.

Among the four decomposed effects, 
the rise in assets has greatly increased all the inequality measures.
Specifically, the rise in asset holdings helps poor households less than it helps wealthy households. 
It significantly increases inequality by 7.3\% among the lower half of the distribution ($50{:}10$ ratio) while having a relatively smaller, yet still significant, 4.6\% increase on the upper half ($90{:}50$ ratio). 
The rise in asset holdings has led to a significant 6.8\% increase in the $75{:}25$ ratio. 
This evidence shows that the sharp rise in S\&P 500 valuations and real estate prices during COVID widens the inequality between the middle and poor families and the rest of society.
Wealthier households, on average, hold substantially larger asset portfolios and are financially more sustainable in the face of the COVID-19 shock.
Their consumption has experienced more growth compared to that of the poor.

Following \cite{ChernozhukovEtAl2013DR}, to help understand the channel of change of inequality, 
we can consider the linear quantile regression models, $Y = X^{'} \beta(U)$, where $U\iid \UnifDist(0,1)$, $X$ is independent of $U$, and $\beta(\cdot)$ maps the unobserved $U$ to a coefficient vector. 
The variance of the outcome variable can be decomposed as
\begin{equation*}
\mathrm{Var}(Y) = 
\underbrace{\mathbb{E}[\beta(U)]' \, \mathrm{Var}[X] \, \mathbb{E}[\beta(U)]}_{\text{between-group inequality}} 
+ 
\underbrace{\operatorname{trace} \left\{ \mathbb{E}[XX'] \, \mathrm{Var}[\beta(U)] \right\}}_{\text{within-group inequality}} .
\end{equation*}
If holding the coefficients constant, changes in $X$ affect the outcome's variance through two channels.  
The first channel is through the between-group inequality and 
reflects how the variance of observed characteristics \(\mathrm{Var}[X]\) is response for the change of inequality.
The second channel is through the within-group inequality, 
which arises from changes in the proportions of $X$ values (characteristic groups) with heterogeneous slopes.

As shown in \cref{tab:emp:male:4factor},
the rise in asset holdings has partially counteracted the net decline in consumption inequality. 
Since the variance of log asset holdings stays similar across the two years, 
the effects of changes in log asset changes through the first channel are minimal. 
This indicates that the change of log assets between 2018 and 2022 affects the log consumption more through the second channel. 
The distribution of log assets has experienced approximately a location shift to the right, and households with higher asset holdings tend to have higher consumption volatility.

In \cref{tab:emp:male:4factor}, the increase in college education attainments between the two years has increased the consumption inequality 
by 1.3\% for the lower half of the distribution ($50{:}10$ ratio) and slightly decreased the $90{:}50$ ratio by 0.1\% among the upper half. 
These effects are obtained by comparing the two counterfactual log consumption distributions: both holding assets in 2018, other characteristics in 2022, and the conditional log consumption distribution in 2022, with only the college education changing from its 2018 distribution to the 2022 distribution.  
The changes in consumption inequality ratios are economically close to zero and statistically insignificant.

Changes in the consumption structure, i.e., the conditional distribution of log consumption, mainly help reduce the inequality gap across the distribution. 
It measures the log consumption distribution differences we would observe if all the explanatory variables (asset holdings, education, and household characteristics) remained at their 2018 joint distribution, with the only change in the conditional distribution of consumption given the explanatory variables.
The $90{:}50$ ratio shrinks twice as much as the $50{:}10$ ratio does ($-2.9$\% vs $-1.4$\%).  
Although these effects are of economically significant magnitude, they are not statistically significant, as the standard errors are large.

The changes in the remaining household characteristics altogether have reduced the inequality considerably. 
These household characteristics include family types (a vector of binary variables indicating married-couple/single-head households, the number and ages of kids), work status, age, race, and whether the household is living in an urban or rural area.
Among them, some family characteristics have experienced significant changes as reported in \cref{tab:stat:male:2}.
For example, a larger proportion of male-headed households are single-person family types. 
The share of families with young kids has shrunk significantly. 
These changes in family type shares may be due to people in their 20s-30s delaying marriage and/or delaying the first baby during COVID.
These self-adjusted behaviors play a role in reducing the consumption inequality over this period. 
These effects are mostly statistically significant, especially for the 90:10, 90:50, and 75:25 ratios.
Such family characteristics changes reduce inequality more on the lower half of the distribution than on the upper half ($-8.0$\% vs $-5.3$\%).

\begin{figure}[htbp]
\centering
\scriptsize
\includegraphics[width=0.95\textwidth]{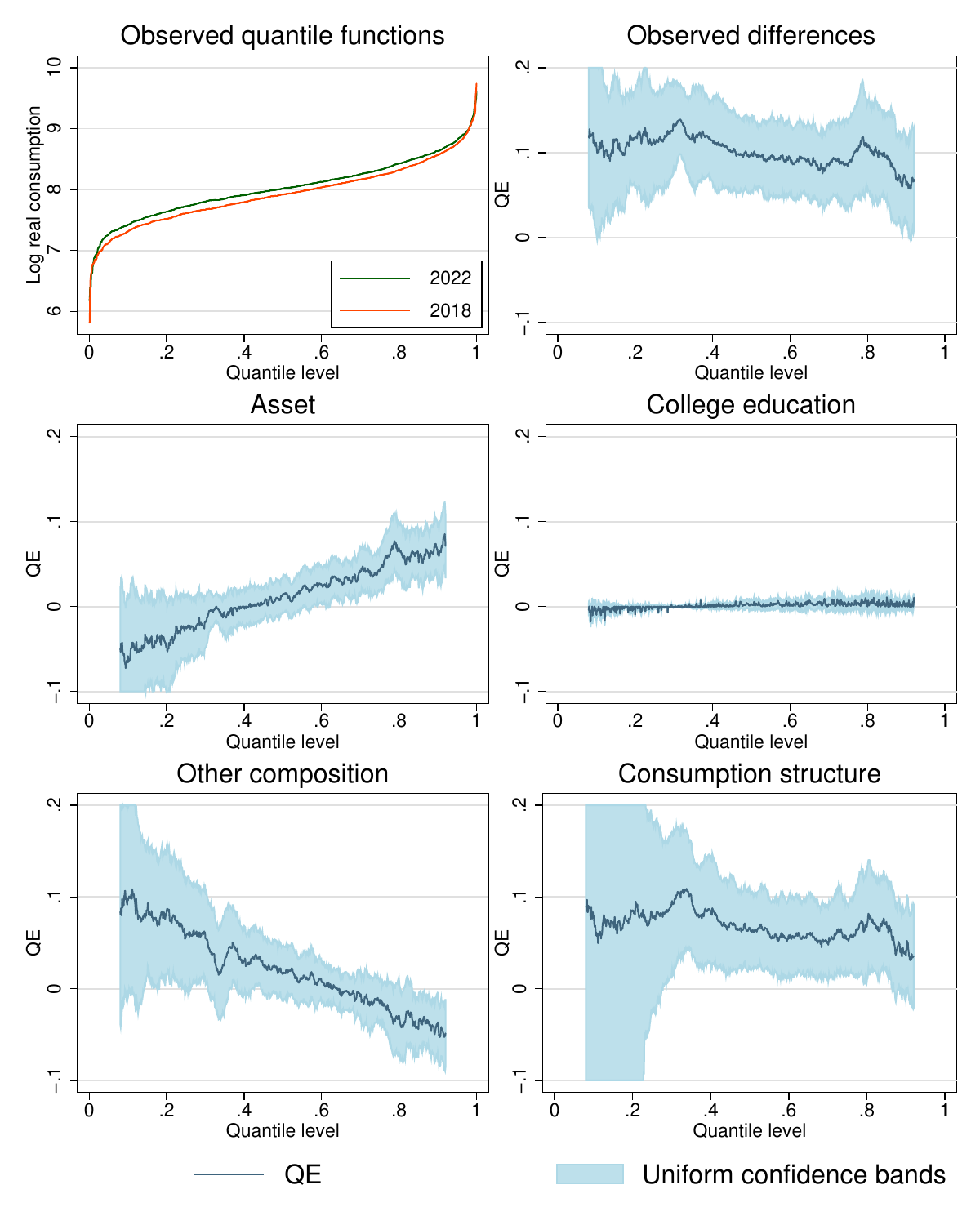}
\caption{Observed quantile functions (top-left), observed differences between the quantile functions (top-right), and their decomposition into four quantile effects (middle and bottom). The 95\% uniform confidence bands are obtained from 500 bootstrap replications following Algorithm 3 in \cite{ChernozhukovEtAl2013DR}.}
\label{fig:male:QE}
\end{figure}

\begin{figure}[htbp]
\centering
\scriptsize
\includegraphics[width=0.95\textwidth]{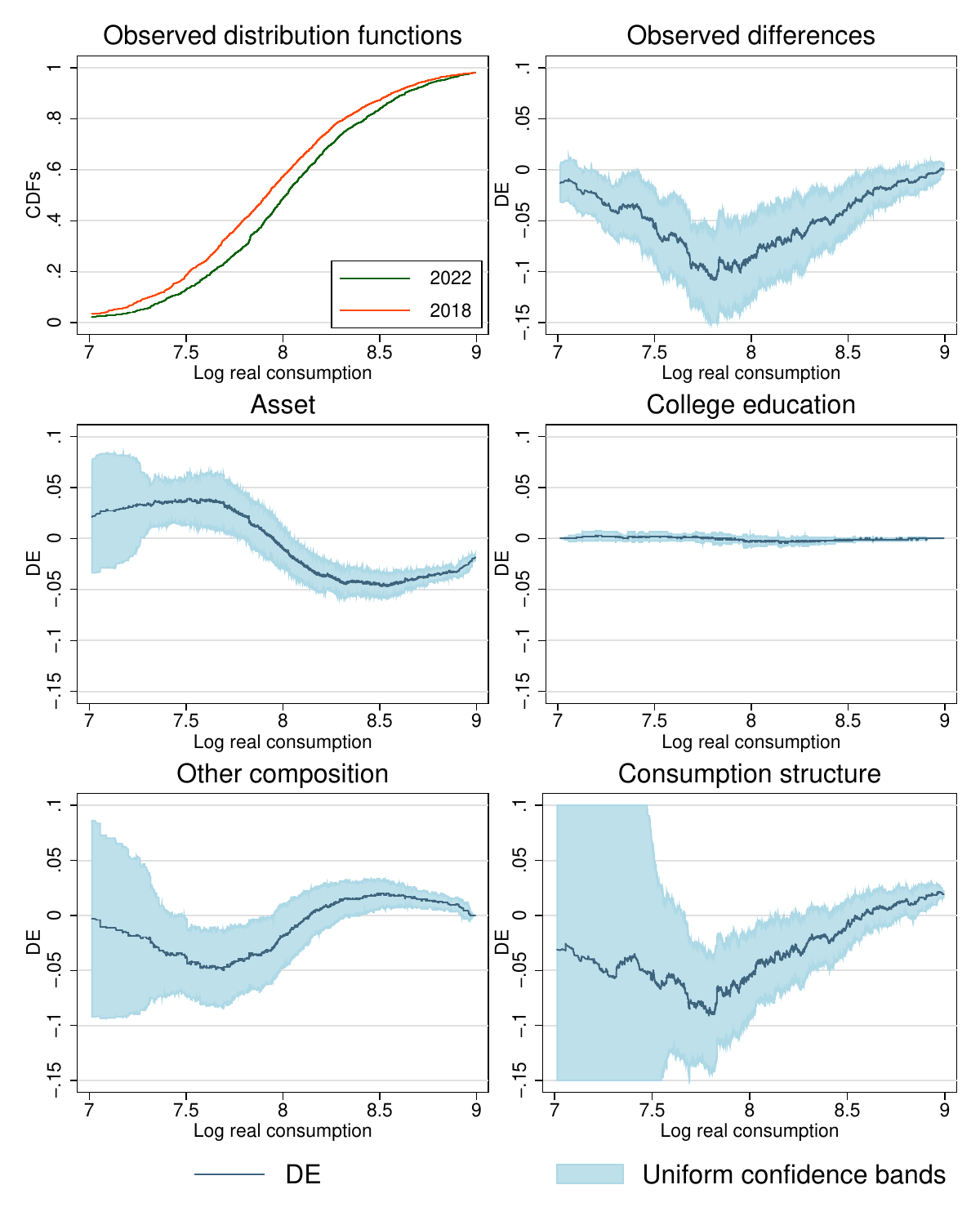}
\caption{Observed distribution functions (top-left), observed differences between the distribution functions (top-right), and their decomposition into four distributional effects (middle and bottom). The 95\% uniform confidence bands are obtained from 500 bootstrap replications following Algorithm 3 in \cite{ChernozhukovEtAl2013DR}.}
\label{fig:male:DE}
\end{figure}

\Cref{fig:male:QE} plots the observed quantile functions in the years 2018 and 2022,
the observed differences between the quantile functions, 
and the decompositions into four quantile effects 
as presented in \cref{eqn:QE:decom}. 
\cref{fig:male:QE} shows that the increase in assets benefits the richer households more and helps less for the households at the bottom of the distribution; on the contrary, the changes in other compositions have increased the consumption for the socially disadvantaged households and slightly decreased it for the households on the upper end of the distribution.
Given the assets, education, and other household characteristics, people consume more in 2022 than in 2018.
Households at the lower quantile of the conditional distribution of consumption increased their consumption more from 2018 to 2022 than those at higher quantile levels, thereby decreasing inequality ratios, such as the 90:10 ratio.
The increase in college education has slightly benefited households at the upper end of the distribution and slightly hurt those at the lower end.

\Cref{fig:male:DE} plots the two years' observed distribution functions, their observed difference, and the decompositions into four distributional effects. 
The results in \cref{fig:male:DE} correspond to those in \cref{fig:male:QE}.

Both \cref{fig:male:QE} and \Cref{fig:male:DE} indicate that there is a large uncertainty about the effect of changes in consumption structure at the lower end of the distribution (below 0.2-quantile levels and correspondingly below the log consumption value of 7.5). 
The distribution regression method is well known for the imprecise estimation at the two tails. 
This is due to the sparsity of data at the two tails, which makes it very noisy to estimate the conditional distribution function at threshold values below which, at the lower tail (or above which, at the upper tail), few observations are available.
For the quantile effects in \cref{fig:male:QE}, we focus on the quantile levels in the (0.08, 0.92) range, which captures the 90:10 ratio and other important inequality measures.
For the distributional effects in \cref{fig:male:DE}, we focus on the (7, 9) range of log consumption, which corresponds approximately to the 0.01 and 0.97 quantiles of the log consumption distribution, respectively. 
Although the sparsity issue also applies to the upper tail, where the log real consumption value is above 9, and the corresponding quantile level is above 0.97,  those households at the extreme upper tail (the top 3\% of the population) are generally not the focus of consumption inequality studies.

We present the empirical results of using the traditional distributional decomposition method in \cref{subsec:MellyResult} and leave the discussions and comparisons there.

The results for female-headed households are presented in \cref{sec:female}.
Compared to male-headed households, 
female-headed households exhibit an increase in the 90:50 ratio and a larger decrease in the 50:10 ratio.
They have experienced a smaller increase in assets over the two years, and thus a smaller magnitude of increase in consumption inequality due to asset changes.
Unlike male-headed households, the consumption structure change for female-headed households has increased consumption inequality slightly, although not significantly.
As in male-headed households, the changes in other household characteristics altogether have decreased consumption inequality for female-headed households.

\section{Conclusion}
\label{sec:conclusion}

We study the effect of the COVID-19 pandemic on changes in U.S. consumption inequality from 2018 to 2022. 
Using \citeposs{ChernozhukovEtAl2013DR} distribution regression method, we can disentangle the effects of our economic variables of interest on changes in consumption inequality.

We find that consumption inequality has declined slightly in the lower half of the distribution and modestly declined in the upper half, resulting in an overall reduction among male-headed households. 
We find that three components have mainly explained the observed changes in inequality, moving in opposite directions: changes in assets, the consumption structure, and all other characteristics. 
The rise in asset holdings has significantly increased consumption inequality across all measures, especially among households at the middle and bottom of the population, partially counteracting the net decline in consumption inequality. 
The increase in the population share of college education has slightly increased the consumption inequality, though not statistically significant. 
Changes in the remaining household characteristics have considerably reduced all the consumption inequality measures.

Future work involves investigating the appropriate economic factors to make the decomposition have causal meaning as discussed in \citet[Section 2.3]{ChernozhukovEtAl2013DR}, and then 
which economic factors contribute to both consumption inequality and income inequality, and add to the literature on whether and how consumption inequality mirrors income inequality.

\Supplemental{\newpage}{%
% \singlespacing
\bibliographystyle{chicago}
\bibliography{_bib}
}

% \onehalfspacing

\appendix

% \paperspacing

\numberwithin{equation}{section}
\numberwithin{theorem}{section}
\numberwithin{assumption}{section}

\pagebreak

\section{Consumption model}
\label{sec:cons}

\subsection{Meyer and Sullivan (2023) consumption model}
\label{secsub:ConsModel}

We briefly review the consumption model proposed in \cite{MeyerSullivan2023}.
Their equation (1) considers the following consumption model with measurement error
\begin{equation}\label{eqn:cons}
    \ln(X_{hjt})=\ln(X_{hjt}^*)+\psi_{t}^j+v_{hjt} ,
\end{equation}
where $X_{hjt}$ denotes the observed consumption of good $j$ for household $h$ at year $t$, 
$X_{hjt}^*$ is the corresponding true latent consumption,  
$\psi_{t}^j$ is a systematic error for good $j$ at year $t$, 
and $v_{hjt}$ is an idiosyncratic error that is assumed to be uncorrelated with the true latent consumption. 
It considers two types of goods $j\in\{w,n\}$: the well-measured category ($j=w$) and not well-measured category ($j=n$).
In other words, \cref{eqn:cons} holds for both $j=w$ and $j=n$.

\cite{MeyerSullivan2023} argue that both the well-measured category and the not-well-measured category have measurement error and reporting issues, i.e., there is a discrepancy between the observed consumption and true latent consumption. 
They assume that the measurement error for the well-measured category is relatively stable over time, 
i.e., $v_{hwt}$ has the same distribution over time; and that the distribution of the measurement error for the not-well-measured category (i.e., $v_{hnt}$) varies arbitrarily over time.

With additional assumptions, 
\cite{MeyerSullivan2023} derive a relationship between the observed well-measured consumption and the latent total consumption $X_{hjt}=X_{ht}^*e^{\alpha+\psi_{t}^w}e^{\epsilon_{hwt}+v_{hwt}}$ in their equation (3), where $X_{ht}^*$ and $\epsilon_{hwt}$ are presented just below in \cref{secsub:ConsAssp}.  
They show that, under the assumption that the error terms of well-measured consumption components are well-behaved as discussed above, changes in observed well-measured consumption can well approximate changes in the latent total consumption.

\subsection{Assumptions in Meyer and Sullivan (2023) consumption model}
\label{secsub:ConsAssp}

The assumptions for \citeposs{MeyerSullivan2023} methods to well approximate the true consumption inequality (of the true latent total consumption) using the well-measured components are: 
(i) The elasticity of the true latent total consumption ($X_{ht}^*=\sum_{j\in\{w,n\} }X_{hjt}^*$) relative to the well-measured components ($X_{hwt}^*$) (minus the idiosyncratic error and systematic error in each good) equals 1. That is, their equation (2) assumes $\ln(X_{hwt}^*)=\alpha + \ln(X_{ht}^*)+\epsilon_{hwt}$, where $\epsilon_{hwt}$ is a well-behaved error term that is independent of the latent total consumption and follows the same distribution over time. 
(ii) The ratio of the weighted average price (i.e., CPI) of the well-measured components bundle relative to the CPI of the total consumption basket stays stable over time. 
(iii) The idiosyncratic error for the well-measured component (i.e., $v_{hwt}$) has the same distribution over time, and this error is independent of the true latent total consumption.

The assumptions (i) and (ii) are testable. 
Following \cite{MeyerSullivan2023}, we examine these two assumptions using our 2018 and 2022 data in \cref{subsec:ConsAsspVerify}.
The assumption (iii) is a restriction on the error term of the well-measured component in their consumption model, which \citet[p.\ 253]{MeyerSullivan2023} argue that it is standard to ignore this error in the inequality literature.

\subsection{Examine Meyer and Sullivan (2023) assumptions for our data}
\label{subsec:ConsAsspVerify}

\begin{table}[htbp]
\centering
\def\sym#1{\ifmmode^{#1}\else\(^{#1}\)\fi}
\caption{Total consumption elasticities of well-measured consumption}
\label{tab:wellmeasured_elasticities}
\begin{threeparttable}
\begin{tabular}{lcc}
\toprule
\multicolumn{3}{c}{Dependent variable: log well-measured consumption}\\
\midrule
& 2018 & 2022 \\
\cmidrule(lr){2-3}
\addlinespace[6pt]
\quad Log total consumption
& \num{0.9368293}\sym{***} & \num{0.9141914}\sym{***} \\
& (\num{0.006551}) & (\num{0.0076896}) \\
\midrule
Number of observations & 1{,}821 & 1{,}689 \\
\bottomrule
\end{tabular}
\begin{tablenotes}[para,flushleft]
\footnotesize
Male-headed households only.
Instrumental variable: log family income. 
Standard errors are in parentheses.
\textsuperscript{***}\,$p<0.01$, \textsuperscript{**}\,$p<0.05$, \textsuperscript{*}\,$p<0.1$.
\end{tablenotes}
\end{threeparttable}
\end{table}

\Cref{tab:wellmeasured_elasticities} reports the coefficients of regressing log well-measured consumption on the log total consumption using log family income as the instrumental variable.
It excludes observations whose household income is in the top or bottom 5 percent of the population.
The estimated elasticities are close to 1, but statistically significant below 1 for the null hypothesis $H_0\colon \textrm{elasticity}=1$.
The magnitudes of the estimates and their statistical significance are similar to those reported in \cite{MeyerSullivan2023}, who find that the estimated coefficients are 0.944 (standard error = 0.001) in 1980 and 0.829 (standard error = 0.009) in 1988.
We tend to think the Assumption (i) above is satisfied for our sample.

\begin{table}[htbp]
\centering
\caption{CPI for all items and well-measured consumption basket }
\label{tab:cpi_weighted}
\begin{threeparttable}
\begin{tabular}{l
S[round-mode=places,round-precision=1,table-format=3.1]
S[round-mode=places,round-precision=1,table-format=3.1]
S[round-mode=places,round-precision=3,table-format=1.3]}
\toprule
Time & {\makecell{ Well-measured basket \\ (1) }}  & {\makecell{ All items basket\\(2) }} & {\makecell{ Ratio\\(1)/(2) }} \\
\midrule
2018:Q4 & 241.441 & 252.052 & 0.958 \\
2022:Q4 & 294.775 & 297.507 & 0.991 \\
\bottomrule
\end{tabular}
\begin{tablenotes}[para,flushleft]
\footnotesize Base year 1982--1984 $=100$.
\end{tablenotes}
\end{threeparttable}
\end{table}

\Cref{tab:cpi_weighted} reports the quarterly CPIs for the all-items basket and the well-measured consumption basket.
The latter is constructed as the weighted average of CPI for each well-measured component.
The final column shows the ratio of the well-measured consumption basket CPI to the all-items basket CPI.
This ratio is above 0.95 and close to 1 in both time periods.

\begin{table}[htbp]
\centering
\caption{Other aggregate price indices}
\label{tab:inflation_indices}
\begin{threeparttable}
\begin{tabular}{l
S[round-mode=places,round-precision=1,table-format=3.1]
S[round-mode=places,round-precision=1,table-format=3.1]
S[round-mode=places,round-precision=1,table-format=3.1]
S[round-mode=places,round-precision=1,table-format=3.1]
S[round-mode=places,round-precision=1,table-format=3.1]
S[round-mode=places,round-precision=1,table-format=3.1]
}
\toprule
Year 
& {\makecell{All-items \\ CPI}}  
& {\makecell{Core \\ CPI}} 
& {\makecell{Supercore \\ CPI}} 
& {PCE} 
& {\makecell{Core \\ PCE}} 
& {\makecell{GDP \\ deflator}} \\
\midrule

2018 
& 251.1 
& 257.6 
& 206.5
& 218.9 
& 221.1
& 215.2\\

2022 
& 292.6
& 294.2
& 230.7
& 249.0 
& 248.5
& 248.3\\

\addlinespace[4pt]
Ratio: 2022/2018 
& {\num[round-mode=places,round-precision=3]{1.165}}
& {\num[round-mode=places,round-precision=3]{1.142}}
& {\num[round-mode=places,round-precision=3]{1.117}}
& {\num[round-mode=places,round-precision=3]{1.137}}
& {\num[round-mode=places,round-precision=3]{1.124}}
& {\num[round-mode=places,round-precision=3]{1.154}} \\

\bottomrule
\end{tabular}

\begin{tablenotes}[para,flushleft]
\footnotesize 1982--1984=100. 
``PCE'' denotes personal consumption expenditures.
\end{tablenotes}
\end{threeparttable}
\end{table}

\Cref{tab:inflation_indices} reports additional aggregate price indices that measure inflation. 
Using a different price index will change the extent to which the real consumption distribution and real asset distribution are scaled, 
which results in a vertical shift of the distributions of log consumptions and log assets.
We use the All-items CPI, which shows the largest increase between the two years, resulting in the smallest gap between the two years' log real consumption distributions compared to other price indices. 
This will affect the gap between the 2018 and 2022 log real consumption distribution at a given quantile level, but not the inequality ratio.  
For example, if we instead use the supercore CPI, 
the change in the log real consumption distribution $Q_\tau(\ln(C_{r,22}))-Q_\tau(\ln(C_{r,18}))$ at a fixed quantile level $\tau$ could be uniformly (over $\tau \in (0,1)$) increased by approximately 4\% ($=100\times(\ln(1.165)-\ln(1.117))$).
In contrast, the 90:10 ratio remains unchanged because of the double differences (between quantiles and between years) of the log distribution.
That is, we can write the gap between the two years' log real consumption as the sum of the gap of the log nominal consumption and a constant reflecting the use of a different price index,
$\ln(C_{\textrm{r},22})-\ln(C_{\textrm{r},18})
= \ln(C_{\textrm{n},22})-\ln(\textrm{CPI}_{22}) - [\ln(C_{\textrm{n},18})-\ln(\textrm{CPI}_{18})]
= \ln(C_{\textrm{n},22})-\ln(C_{\textrm{n},18}) - \ln(\textrm{CPI}_{22}/\textrm{CPI}_{18})$, where $C_r$ denotes real consumption and $C_n$ denotes nominal consumption.
This shows that the price indices matter for the gap of the log real consumption distribution between two years.
However, the change of the inequality ratios, for example, the 90:10 ratio, between the two years is invariant to the choice of price indices,
\begin{align*}
 &   Q_{0.9}(\ln(C_{\textrm{r}, 22})) - Q_{0.1}(\ln(C_{\textrm{r}, 22})) -[Q_{0.9}(\ln(C_{\textrm{r}, 18}))-Q_{0.1}(\ln(C_{\textrm{r}, 18}))] \\
& = Q_{0.9}(\ln(C_{\textrm{n},22})- \ln(\textrm{CPI}_{22}) ) - Q_{0.1}(\ln(C_{\textrm{n},22})-\ln(\textrm{CPI}_{22}))\\ 
& \quad -[Q_{0.9}(\ln(C_{\textrm{n},18})- \ln(\textrm{CPI}_{18}) )-Q_{0.1}(\ln(C_{\textrm{n},18})- \ln(\textrm{CPI}_{18}))]\\
& =  Q_{0.9}(\ln(C_{\textrm{n}, 22})) - Q_{0.1}(\ln(C_{\textrm{n}, 22})) -[Q_{0.9}(\ln(C_{\textrm{n}, 18}))-Q_{0.1}(\ln(C_{\textrm{n}, 18}))]
\end{align*}
where the last equality comes from the fact that $Q_\tau(Y+c)=Q_\tau(Y)+c$ for any random variable $Y$, constant $c$, and quantile level $\tau\in(0,1)$, and the CPI index in each year is a constant.

\section{Imputation of well-measured consumption components}
\label{sec:impute}

\cite{MeyerSullivan2023} consider the following categories as the well-measured consumption: food at home, utilities, gas, imputed rental equivalents of housing, and imputed service flows of all vehicles owned by the household.
We impute the quarterly equivalent values for the housing and vehicle expenditures in the same way as in \cite{MeyerSullivan2023}.

\subsection{Imputation of Vehicle Service Flows}
\label{sec:veh}

Following the same method in \cite{MeyerSullivan2023}, the quarterly vehicle service flow is computed as the product of the (observed or imputed) current market value of the vehicle and the fixed depreciation rate from \cite{MeyerSullivan2023}.
76.4\% of the samples have a missing value for the recorded vehicle purchase price and thus require imputation. 
The imputed vehicle value is included in both consumption and assets, although it accounts for only a small share of each.

For vehicles purchased within 12 months of the interview and for which a purchase price is available (9.5\% of the sample), we use the reported purchase price as the current market value, and no imputation is needed.

For vehicles purchased more than 12 months before the interview and with a purchase price (14.1\% of the sample), we compute the current market value by applying a geometric depreciation based on the average number of years of existing vehicles to the recorded purchase price. And no imputation is needed.

For the remaining 76.4\% of the sample without a recorded purchase price, we impute the current market value. 
We first regress the log purchase price on household and vehicle characteristics using the subsample with a recorded purchase price. 
Then we use the estimated coefficients and the household and vehicle characteristics to impute the log real purchase price and convert it to the predicted purchase price following \citeposs{MeyerSullivan2023} method.

\subsection{Imputation of Rental Equivalents}
\label{sec:ren}

Following \cite{MeyerSullivan2023}, we impute the household quarterly rental equivalents for households that do not have a reported rental equivalent, which accounts for 35.2\% of the total sample. 
The imputed rental equivalents are counted only as a component of consumption, not as an asset. 

We impute monthly rental equivalents and multiply them by three to obtain the quarterly value for housing expenditure.
Specifically, we regress the log monthly rent on household and unit characteristics using the subsample that reports rent payments.
Then we use the estimated coefficients, the household characteristics, and unit characteristics to compute the fitted outcome and take the exponential to impute the monthly rental equivalents.

\section{Traditional distributional decomposition method and results}
\label{sec:compare_melly}

\subsection{Compare to traditional distributional decomposition method}
\label{subsec:MellyModel}

The traditional distributional decomposition method in \cite{Melly2005} attributes the differences between two observed quantile functions to three effects: due to changes in residuals, coefficients, and characteristics 
\begin{align}
\label{eqn:Melly:QE:decom}
    & \underbrace{\hat{Q}_\tau(\vecf{\hat{\beta}}^{22}, \vecf{X}^{22} ) -\hat{Q}_\tau(\vecf{\hat{\beta}}^{18}, \vecf{X}^{18} ) }_{\textrm{observed unconditional quantile function differences}}\notag \\
    & = \underbrace{\hat{Q}_\tau(\vecf{\hat{\beta}}^{22}, \vecf{X}^{22} ) - \hat{Q}_\tau(\vecf{\hat{\beta}}^{m22,r18}, \vecf{X}^{22} )}_{\textrm{QE due to residuals change} } 
        + \underbrace{\hat{Q}_\tau(\vecf{\hat{\beta}}^{m22,r18}, \vecf{X}^{22} ) - \hat{Q}_\tau(\vecf{\hat{\beta}}^{18}, \vecf{X}^{22} ) }_{\textrm{QE due to coefficients change} } \notag \\
    & \quad 
        + \underbrace{\hat{Q}_\tau(\vecf{\hat{\beta}}^{18}, \vecf{X}^{22} ) - \hat{Q}_\tau(\vecf{\hat{\beta}}^{18}, \vecf{X}^{18} ) }_{\textrm{QE due to characteristics change }}, \tau \in (0,1),
\end{align}
where $\vecf{X}=(X_1,X_2,\vecf{X}_3')'$ using our paper's notation is a full vector of explanatory variables.

The $\hat{Q}_\tau(\vecf{\hat{\beta}}^{22}, \vecf{X}^{22})$ and $\hat{Q}_\tau(\vecf{\hat{\beta}}^{18}, \vecf{X}^{18} )$
are the unconditional observed sample quantile functions in the years 2018 and 2022. 
(Although they are observed in data and can be constructed directly from the empirical cumulative distribution function of log consumption in both years, \citeposs{Melly2005} method constructs them under a set of assumptions on the model specifications as described below.)
They are constructed by first using a collection of linear quantile regressions to obtain the estimated conditional function 
\begin{equation}\label{eqn:Melly:step1:CQF}
    \hat{Q}_\tau (Y^t \mid \vecf{X}=\vecf{X}^t) = \vecf{X}^{t\prime} \vecf{\hat{\beta}}^t(\tau), t\in\{18, 22\},
\end{equation}
at a grid of quantile levels $(\tau_0,\tau_1,\ldots,\tau_J)$ (hypothetically $ \tau_0=0$ and $\tau_J=1$)
and then integrate over the grid of $\tau$ and covariates $\vecf{X}$ to solve for the unconditional $\tilde{\tau}$-th quantile function $q_{\tilde{\tau}}$ satisfying 
\begin{equation}\label{eqn:Melly:step1:int}
    \int_{\mathcal{X}_t} \int_0^1 \Ind{\vecf{X}^{t\prime} \vecf{\hat{\beta}}^t(\tau)\leq q_{\tilde{\tau}}} \diff \tau \diff F_{\vecf{X}^t}(\vecf{x}) =\tilde{\tau}, \tilde{\tau}\in (0,1), t\in\{18, 22\}.
\end{equation}
That is, $\hat{Q}_{\tilde{\tau}}(\vecf{\hat{\beta}}^{22}, \vecf{X}^{22})$ and $\hat{Q}_{\tilde{\tau}}(\vecf{\hat{\beta}}^{18}, \vecf{X}^{18} )$ 
are the $q_{\tilde{\tau}}$ solved in \cref{eqn:Melly:step1:int} with $t=18, 22$ respectively.

The $\hat{Q}_\tau(\vecf{\hat{\beta}}^{18}, \vecf{X}^{22} ) $ is the counterfactual unconditional quantile function $q_{\tilde{\tau}}$ solved in \cref{eqn:Melly:step1:int} with the conditional quantile function coefficient in 2018 ($\vecf{\hat{\beta}}^{18}$) and the characteristics in 2022 ($\vecf{X}^{22}$).

The hypothetical slope $\vecf{\hat{\beta}}^{m22,r18}$ is constructed as $\vecf{\hat{\beta}}^{m22,r18}(\tau_j)\equiv \vecf{\hat{\beta}}^{22}(0.5) + \vecf{\hat{\beta}}^{18}(\tau_j) - \vecf{\hat{\beta}}^{18}(0.5)$ at all $\tau_j$s over the grid vector $(\tau_0,\tau_1,\ldots,\tau_J)$, 
where $\vecf{\hat{\beta}}^{22}(0.5)$ and $\vecf{\hat{\beta}}^{18}(0.5)$ are the slope for the median ($0.5$-quantile) conditional quantile function in the two years 
and $\vecf{\hat{\beta}}^{18}(\tau_j)$ is the $\tau_j$-th conditional quantile function slope.
\cite{Melly2005} considers $\vecf{\hat{\beta}}^{18}(\tau_j) - \vecf{\hat{\beta}}^{18}(0.5)$ as the ``residual'' part and use the superscript ``m22, r18'' to denote the slope $\vecf{\hat{\beta}}^{m22,r18}(\tau_j)$ is the median slope in 2022 plus the residual from year 2018. 
Using the constructed slope  $\vecf{\hat{\beta}}^{m22,r18}(\tau_j)$ and the 2022's covariates $\vecf{X}^{22}$, people can solve for the constructed counterfactual $q_{\tilde{\tau}}$ (i.e., $\hat{Q}_\tau(\vecf{\hat{\beta}}^{m22,r18}, \vecf{X}^{22} )$) from in \cref{eqn:Melly:step1:int}.

This traditional distributional decomposition method relies on many strong assumptions that are less likely to hold in practice. 
First, it requires the conditional quantile function to be linear in both coefficients and regressors at all quantile levels. 
These restrictions often encounter the quantile crossing problems 
(i.e., the estimated conditional quantile curves at different quantile levels cross each other, violating the monotonicity  condition) as proposed in \cite{ChernozhukovEtAl2010}.
In addition, this estimation method might suffer from the misspecification error. 
That is, if the true conditional quantile function (CQF) is nonlinear, then running linear quantile regression and getting the estimated slope only allows us to obtain the best linear predictor of the CQF.
The gap between the true CQF and its best linear predictor can be arbitrarily large, in which case the estimated effects and the constructed counterfactual functions, which are essentially based on these linear approximations, are unreliable.

Unlike the linear conditional quantile function assumptions required by \citeposs{Melly2005} method, the distribution regression method allows a very flexible nonlinear model for the CQF, which is often the case in practice.

Second, the construction of \citeposs{Melly2005} ``residual'' components could be arguably arbitrary, 
because it essentially compares the linear conditional quantile slope at all quantiles to the median slope.
That is, the ``residual component'' in the traditional method is determined by where the conditional median of the outcome is, or more specifically, the (potentially misspecified) conditional median slope.

\subsection{Decomposition results using the traditional method}
\label{subsec:MellyResult}

\begin{table}[htbp]
\centering
\caption{\label{tab:emp:male:melly}Decomposition of consumption inequality changes by Melly (2005) method}
\sisetup{round-precision=1,round-mode=places,detect-all} 
\begin{threeparttable}
\begin{tabular}{l
S[table-format=-2.1]
S[table-format=-2.1]
S[table-format=-2.1]
S[table-format=-2.1]
S[table-format=-2.1]
S[table-format=-2.1]
S[table-format=-2.1]}
\toprule
 & &&  \multicolumn{3}{c}{Effects of } \\
\cmidrule{4-6}
{Statistics} & {Total Changes} && {Coefficients} & {Characteristics} & {Residuals}  \\
% {\makecell{Total\\Changes}} 
\midrule
  SD       &   -1.492478 &&    -2.307105  &   0.309408 & 0.505219 \\
         %&             &&   154.59      &  -20.73    & -33.86    \\
  90--10   &   -5.387038 && -5.357798  &   -0.172255 &   0.143015 \\
  %& &&  -2.654798 &  99.457216 &   3.197583 \\
  50--10   &   -1.742804 && -1.510127  &   -0.473924 &   0.241247 \\
  %& && -13.842463 &  86.649273 &  27.193190 \\
  90--50   &   -3.644234 && -3.847671  &    0.301669 &  -0.098231  \\
  %& &&   2.695518 & 105.582435 &  -8.277981 \\
  75--25   &   -4.822392 && -2.909181  &   -0.503669 &  -1.409542  \\
  %& &&  29.229105 &  60.326514 &  10.444381 \\
  95--5    &   -4.459799 && -7.067763  &    1.600629 &  1.007334 \\
  %& && -22.586982 & 158.477165 & -35.890160 \\
  Gini     &  -0.167616  &&   -0.173667  &   -0.006397 & 0.012449   \\
         %&             &&   103.61      &   3.82      & -7.43       \\
\bottomrule
\end{tabular}
\begin{tablenotes}[para,flushleft]
\footnotesize{}
Male-headed households only. 
All values are in percentages. 
\end{tablenotes}
\end{threeparttable}
\end{table}

\Cref{tab:emp:male:melly} reports the decomposition results using \citeposs{Melly2005} method. 
The total observed changes of consumption inequality statistics 
are slightly different from those reported in \cref{tab:emp:male:4factor}.
% \cref{tab:emp:male:6factor}. 
This reflects that the restrictions of linear CQFs at all quantile levels, which \citeposs{Melly2005} method has imposed on constructing the observed unconditional quantile functions for the two years.

\citeposs{Melly2005} method decomposes the total changes into three effects: those explained by changes in coefficients, characteristics, and unexplained parts summarized in residuals. 
\Cref{tab:emp:male:melly} shows that most of the observed inequality changes are explained by the overall coefficient changes, which have shrunk the inequality throughout the distribution.
The overall changes in characteristics have slightly decreased consumption inequality among the lower half of the distribution (50:10 ratio) and slightly increased the inequality in the upper half (90:50 ratio).
The unexplained parts in residuals are relatively small in magnitude.
As mentioned earlier, this traditional method is not very informative and cannot break down changes in consumption inequality into specific changes in explanatory variables.

\section{Additional effects of married-no-kids and work status}
\label{sec:emp:extra}

In this section, we additionally separate out the effects of a family type (married couple with no kids) and work status from the rest of the household characteristics following the results in \cref{sec:emp}.

\begin{table}[htbp]
\centering
\caption{\label{tab:emp:male:6factor}Extended decomposition of consumption inequality changes with additional factors}
\sisetup{round-precision=1,round-mode=places,detect-all}
\begin{threeparttable}
\begin{tabular}{lccccccc}
\toprule
& & \multicolumn{6}{c}{Effects of changes in} \\
\cmidrule{3-8}
{Statistics} 
  & {\makecell{Total\\Changes}} 
  & {Assets} 
  & {\makecell{College\\Education}} 
  & {\makecell{Married\\No Kids}} 
  & {\makecell{Work\\ Status}} 
  & {\makecell{Remaining\\Covariates}} 
  & {\makecell{Consumption\\Structure}} \\
\midrule

SD          
  & \num{-1.544154} (\num{1.499138})
  & \num{4.375745} (\num{2.055125})
  & \num{0.298745} (\num{0.222765})
  & \num{0.359215} (\num{0.252663})
  & \num{-0.014895} (\num{0.140873})
  & \num{-5.026509} (\num{3.500409})
  & \num{-1.536455} (\num{10.028599}) 
\\[0.65em]

% Share estimates: Total=.; Assets=-283.37493; Education=-19.346865; Family=-23.262889; Work=0.964617; Remaining=325.51863; Structure=99.501429
% Share SEs: Total=.; Assets=437.0776;  Education=24.10748;   Family=24.31228;   Work=12.672881; Remaining=486.01939; Structure=540.14785

90--10      
  & \num{-4.490374} (\num{5.543086})
  & \num{11.923610} (\num{4.212275})
  & \num{1.173945} (\num{0.673783})
  & \num{0.639751} (\num{0.970771})
  & \num{0.192725} (\num{0.131271})
  & \num{-14.137197} (\num{5.549463})
  & \num{-4.283209} (\num{39.091794})
\\[0.65em]

% Share estimates: Total=.; Assets=-265.53714; Education=-26.143597; Family=-14.247169; Work=-4.291954; Remaining=314.8334; Structure=95.386459
% Share SEs: Total=.; Assets=567.48181; Education=10.992001;  Family=15.411321;  Work=2.548594;  Remaining=364.55171; Structure=371.87821

50--10      
  & \num{-0.801027} (\num{4.764475})
  & \num{7.318913} (\num{3.348183})
  & \num{1.283847} (\num{0.683176})
  & \num{0.421967} (\num{0.891670})
  & \num{0.032788} (\num{0.081855})
  & \num{-8.491578} (\num{4.858571})
  & \num{-1.366965} (\num{7.922596})
\\[0.65em]

% Share estimates: Total=.; Assets=-913.6908;  Education=-160.27513; Family=-52.67826; Work=-4.09323; Remaining=1060.0859; Structure=170.65148
% Share SEs: Total=.; Assets=406.48232; Education=18.988753;  Family=22.567902; Work=1.822992; Remaining=399.91413; Structure=304.45722

90--50      
  & \num{-3.689347} (\num{2.630227})
  & \num{4.604698} (\num{1.772209})
  & \num{-0.109902} (\num{0.344584})
  & \num{0.217784} (\num{0.295228})
  & \num{0.159937} (\num{0.202090})
  & \num{-5.645619} (\num{1.742899})
  & \num{-2.916244} (\num{3.047758})
\\[0.65em]

% Share estimates: Total=.; Assets=-124.81066; Education=2.978908; Family=-5.903048; Work=-4.3351; Remaining=153.0249; Structure=79.044992
% Share SEs: Total=.; Assets=167.94603; Education=10.619623; Family=10.890222; Work=5.943182; Remaining=163.20282; Structure=61.660091

75--25      
  & \num{-2.744663} (\num{2.997578})
  & \num{6.792994} (\num{2.020490})
  & \num{0.435589} (\num{0.388639})
  & \num{0.274344} (\num{0.690303})
  & \num{0.000000} (\num{0.163438})
  & \num{-7.682331} (\num{2.979675})
  & \num{-2.565259} (\num{5.013110})
\\[0.65em]

% Share estimates: Total=.; Assets=-247.49829; Education=-15.870415; Family=-9.995539; Work=0; Remaining=279.90072; Structure=93.463524
% Share SEs: Total=.; Assets=379.84419; Education=14.513418;  Family=22.82106;   Work=5.162587; Remaining=339.49094; Structure=119.4104

95--5       
  & \num{-7.520209} (\num{7.205908})
  & \num{13.112467} (\num{7.118399})
  & \num{0.921580} (\num{0.713682})
  & \num{1.085600} (\num{0.972457})
  & \num{0.000000} (\num{0.038318})
  & \num{-13.548963} (\num{14.146978})
  & \num{-9.090894} (\num{32.054363})
\\[0.65em]

% Share estimates: Total=.; Assets=-174.36307; Education=-12.254715; Family=-14.435762; Work=0; Remaining=180.16736; Structure=120.88618
% Share SEs: Total=.; Assets=372.50263; Education=7.07796;    Family=11.204987;  Work=0.831778; Remaining=297.50488; Structure=324.31608

Gini        
  & \num{-0.424456} (\num{0.304382})
  & \num{0.928373} (\num{0.345731})
  & \num{0.064096} (\num{0.046938})
  & \num{0.084711} (\num{0.051492})
  & \num{-0.004243} (\num{0.030289})
  & \num{-1.111303} (\num{0.593249})
  & \num{-0.386090} (\num{1.570942})
\\

% Share estimates: Total=.; Assets=-218.72074; Education=-15.100769; Family=-19.95759; Work=0.999607; Remaining=261.81838; Structure=90.961119
% Share SEs: Total=.; Assets=211.07714; Education=16.505974;  Family=21.73879;  Work=8.756487; Remaining=344.72785; Structure=337.35556

\bottomrule
\end{tabular}
\begin{tablenotes}[para,flushleft]
\footnotesize{}
Male-headed households only. 
All values are in percentages.  
Bootstrapped standard errors with 500 repetitions in parentheses. 
\end{tablenotes}
\end{threeparttable}
\end{table}

In \cref{tab:emp:male:6factor}, the total observed changes, the effects of changes in assets, college education, and the consumption structure remain the same as in \cref{tab:emp:male:4factor}.
The sum of the effects of changes in family structure, work status, and remaining covariates in \cref{tab:emp:male:6factor} equals the effect of changes in remaining covariates in \cref{tab:emp:male:4factor}.

We consider a specific family type: a married couple with no kids.
This category has experienced a significant 3 percentage points drop between the two years.
This decline in the married-couple-with-no-kids group has increased consumption inequality by 0.4\% for the lower half of the distribution ($50{:}10$ ratio) and by 0.2\% for the upper half ($90{:}50$ ratio). 
It may be because poor households choose to delay getting married or having their first baby during COVID in response to financial and job insecurity. 
Additionally, this family type's change has increased the $75{:}25$ consumption ratio by 0.3\% and the $95{:}5$ ratio by 1.1\%. 
The drop in this family type's share has only slightly increased the inequality measures, despite none of these effects being statistically significant.

As presented earlier in \cref{tab:stat:male:2}, the portion of the population who do not have work remains the same for the two years. 
Therefore, as expected, the inequality changes attributable to the changes in work status, if any, are also minimal or zero.

After separating out the effects of changes in one family type and work status, the changes in the remaining characteristics have significantly decreased the consumption inequality, whose magnitude and patterns are similar to those reported in \cref{tab:emp:male:4factor}.

\begin{figure}[htbp]
\centering
\scriptsize
\includegraphics[width=\textwidth]{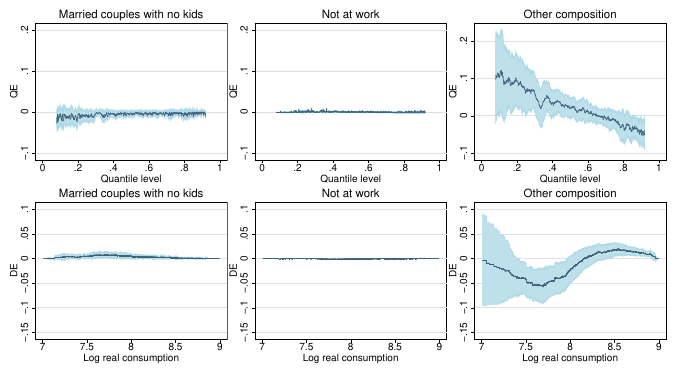}
\caption{Additional decompositions of QE and DE with 95\% uniform confidence bands.}
\label{fig:male:sixfactor:QE:DE}
\end{figure}

\Cref{fig:male:sixfactor:QE:DE} plots the quantile effects and distributional effects when additionally separating out the family type and work status from the rest of the household characteristics.

\section{Robustness check on decomposition sequences}
\label{sec:alt:decom}

\begin{table}[htbp]
\centering
\caption{\label{tab:emp:sixfactor:alt}An alternate decomposition sequence of consumption inequality changes}
\sisetup{round-precision=1,round-mode=places,detect-all}
\begin{threeparttable}
\begin{tabular}{lccccccc}
\toprule
& & \multicolumn{6}{c}{Effects of changes in} \\
\cmidrule{3-8}
{Statistics} 
  & {\makecell{Total\\Changes}} 
  & {\makecell{College\\Education}} 
  & {Assets} 
  & {\makecell{Work\\Status}} 
  & {\makecell{Married\\No Kids}} 
  & {\makecell{Remaining\\Covariates}} 
  & {\makecell{Consumption\\Structure}} \\
\midrule

SD          
  & \num{-1.544154} (\num{1.508694})
  & \num{0.200197}  (\num{8.757811})
  & \num{4.678648}  (\num{1.905368})
  & \num{-0.007036} (\num{0.145617})
  & \num{0.317463}  (\num{0.236352})
  & \num{-5.196970} (\num{3.329686})
  & \num{-1.536455} (\num{9.347917}) \\[0.65em]
  
% Share estimates: Total=.; Edu=-12.964825; Assets=-302.99101; Work=0.455652; Family=-20.559033; Remaining=336.55779; Structure=99.501429
% Share SEs: Total=.; Edu=384.091; Assets=431.08591; Work=11.576395; Family=25.532994; Remaining=506.56058; Structure=518.97382

90--10      
  & \num{-4.490374} (\num{5.564329})
  & \num{0.000000}  (\num{38.277731})
  & \num{13.813135} (\num{4.770071})
  & \num{0.000000}  (\num{0.199564})
  & \num{0.577521}  (\num{0.914421})
  & \num{-14.597820} (\num{5.585869})
  & \num{-4.283209} (\num{38.298347}) \\[0.65em]
  
% Share estimates: Total=.; Edu=0; Assets=-307.61659; Work=0; Family=-12.861307; Remaining=325.09143; Structure=95.386459
% Share SEs: Total=.; Edu=26.590446; Assets=326.95255; Work=3.024163; Family=17.445605; Remaining=371.41962; Structure=348.63808

50--10      
  & \num{-0.801027} (\num{4.788029})
  & \num{0.385591}  (\num{1.442297})
  & \num{8.472124}  (\num{4.196002})
  & \num{0.000000}  (\num{0.092530})
  & \num{0.167012}  (\num{0.902156})
  & \num{-8.458790} (\num{5.128811})
  & \num{-1.366965} (\num{7.631522}) \\[0.65em]
  
% Share estimates: Total=.; Edu=-48.137049; Assets=-1057.6574; Work=0; Family=-20.849738; Remaining=1055.9927; Structure=170.65148
% Share SEs: Total=.; Edu=23.823919; Assets=343.95033; Work=2.384416; Family=26.025922; Remaining=429.32533; Structure=237.09162

90--50      
  & \num{-3.689347} (\num{2.615672})
  & \num{-0.385591} (\num{0.201611})
  & \num{5.341010}  (\num{1.821238})
  & \num{0.000000}  (\num{0.218392})
  & \num{0.410509}  (\num{0.320577})
  & \num{-6.139031} (\num{1.958348})
  & \num{-2.916244} (\num{2.957100}) \\[0.65em]
  
% Share estimates: Total=.; Edu=10.451469; Assets=-144.76846; Work=0; Family=-11.126866; Remaining=166.39887; Structure=79.044992
% Share SEs: Total=.; Edu=6.356418; Assets=168.76896; Work=6.546032; Family=11.021893; Remaining=170.18886; Structure=61.244815

75--25      
  & \num{-2.744663} (\num{2.997578})
  & \num{0.406305}  (\num{0.390114})
  & \num{6.965635}  (\num{2.272249})
  & \num{0.000000}  (\num{0.163438})
  & \num{0.437987}  (\num{0.670175})
  & \num{-7.989332} (\num{3.038017})
  & \num{-2.565259} (\num{4.988475}) \\[0.65em]
  
% Share estimates: Total=.; Edu=-14.803461; Assets=-253.78838; Work=0; Family=-15.957787; Remaining=291.08611; Structure=93.463524
% Share SEs: Total=.; Edu=15.509962; Assets=254.28159; Work=4.137102; Family=26.184233; Remaining=339.91287; Structure=118.29161

95--5       
  & \num{-7.520209} (\num{7.277345})
  & \num{0.934321}  (\num{30.142920})
  & \num{14.013079} (\num{6.857338})
  & \num{0.304581}  (\num{0.120286})
  & \num{0.555440}  (\num{1.007737})
  & \num{-14.236736} (\num{18.536780})
  & \num{-9.090894} (\num{31.849116}) \\[0.65em]
  
% Share estimates: Total=.; Edu=-12.42414; Assets=-186.33896; Work=-4.05016; Family=-7.385964; Remaining=189.31304; Structure=120.88618
% Share SEs: Total=.; Edu=239.20728; Assets=247.96054; Work=1.188245; Family=11.246083; Remaining=319.78283; Structure=303.90511

Gini        
  & \num{-0.424456} (\num{0.308012})
  & \num{0.035876}  (\num{1.351885})
  & \num{1.001990}  (\num{0.408032})
  & \num{-0.001820} (\num{0.029487})
  & \num{0.075750}  (\num{0.052442})
  & \num{-1.150162} (\num{0.601912})
  & \num{-0.386090} (\num{1.456681}) \\
  
% Share estimates: Total=.; Edu=-8.452166; Assets=-236.06463; Work=0.428817; Family=-17.846454; Remaining=270.97331; Structure=90.961119
% Share SEs: Total=.; Edu=275.04564; Assets=302.65858; Work=7.978795; Family=20.222002; Remaining=361.68593; Structure=305.04047

\bottomrule
\end{tabular}
\begin{tablenotes}[para,flushleft]
\footnotesize{}
Male-headed households only.  
All values are in percentages.  
Bootstrapped standard errors with 500 repetitions in parentheses.  
\end{tablenotes}
\end{threeparttable}
\end{table}

In this section, we conduct a robustness check to see whether and to what extent the results are sensitive to the decomposition sequence.
That is, we decompose the change in consumption inequality in a different order of the explanatory factors. 
Specifically, we decompose using the sequence: 
$X_1=\textrm{College education}$, $X_2=\textrm{Assets}$, $X_3=\textrm{Work status}$, $X_4=\textrm{Married no kids}$, 
and $\vecf{X}_5=\textrm{all the remaining variables},$ with results presented in \Cref{tab:emp:sixfactor:alt}.

Each decomposition sequence is associated with a specific economic meaning for the decomposed effects.
For example, the effects due to the change in assets in \cref{tab:emp:sixfactor:alt} measure the differences between two counterfactual log consumption distributions when assets change from 2018's distribution to 2022 distribution, holding education in 2018 distribution, all other variables, and the conditional log consumption distribution in 2022.
For comparison, the effects due to assets in \cref{tab:emp:male:6factor} measure the differences between two counterfactual log consumption distributions when assets change from the 2018 to 2022 distribution, holding all other variables (including education) and the conditional consumption distribution in 2022. 
From the economic meaning, these two effects will not be identical, although they should be very similar.

The estimates in \cref{tab:emp:sixfactor:alt} exhibit a similar pattern and magnitude as in \cref{tab:emp:male:6factor}.
Changes in asset holdings have significantly increased the consumption inequality, especially for the lower half of the distribution. 
Changes in college education and in the family type of married couples with no kids have increased the inequality a bit, although not statistically significant. 
The effects due to changes in work status are mostly near zero. 
This evidence indicates that the decomposition sequence does not qualitatively change the results pattern, although the numerical differences reflect the specific economic meanings of that decomposition sequence. 
In practice, researchers could choose the decomposition sequence based on the economic meanings they are mostly interested in.

\section{Female-headed households results}
\label{sec:female}

\subsection{Summary statistics for female-headed households}
\label{sec:female:stat}

\begin{table}[htbp]
\centering
\sisetup{round-mode=places,round-precision=2,detect-all}
\begin{threeparttable}
\caption{Descriptives of the consumptions and assets for female-headed households}
\label{tab:stat:female:1}
\begin{tabular}{l
S[table-format=3.3]
S[table-format=3.3]
S[table-format=3.3]
S[table-format=5.3]
c}
\toprule
\multicolumn{6}{c}{Year 2018} \\
\midrule
{Variables} & {Mean} & {SD} & {Min} & {Max} & {N}\\
\midrule
Nominal consumption (in \$1000) & 7.120903 & 3.960175 & 1.624985 & 57.132297 & 2188\\
Real consumption (in \$1000)    & 2.835804 & 1.577086 & 0.647128 & 22.752172 & 2188 \\
Log real consumption            & 7.828941 & 0.484748 & 6.472545 & 10.032416 & 2188 \\
Nominal asset holdings (in \$1000) & 31.298179 & 107.090329 & 0.000453 & 2528.980250 & 2188 \\
Real asset holdings (in \$1000)    & 12.464081 & 42.647289 & 0.000180 & 1007.132500 & 2188 \\
Log real asset holdings            & 8.565941 & 1.179271 & -1.712690 & 13.822618 & 2188 \\
\midrule
\multicolumn{6}{c}{Year 2022} \\
\midrule
{Variables} & {Mean} & {SD} & {Min} & {Max} & {N}\\
\midrule
Nominal consumption (in \$1000) & 9.099329 & 4.824438 & 1.756274 & 49.727031  & 1788\\
Real consumption (in \$1000)    & 3.109234 & 1.648507 & 0.600117 & 16.991689  & 1788 \\
Log real consumption            & 7.922168 & 0.488501 & 6.397125 & 9.740479  & 1788 \\
Nominal asset holdings (in \$1000) & 41.309046 & 126.239319 & 0.001000 & 2383.791000  & 1788 \\
Real asset holdings (in \$1000)    & 14.115271 & 43.135883 & 0.000342 & 814.539625  & 1788 \\
Log real asset holdings            & 8.733349 & 1.095308 & -1.073824 & 13.610378  & 1788 \\
\bottomrule
\end{tabular}
\begin{tablenotes}[para,flushleft]
\footnotesize{}
\end{tablenotes}
\end{threeparttable}
\end{table}

\begin{table}[ht]
\centering
\def\sym#1{\ifmmode^{#1}\else\(^{#1}\)\fi}
\sisetup{round-precision=2,round-mode=places,detect-all}
\begin{threeparttable}
\caption{Summary statistics for characteristics of the female-headed household heads}
\label{tab:stat:female:2}
\begin{tabular}{l
S[table-format=3.2,round-precision=2]
S[table-format=3.2,round-precision=2]
c
S[table-format=3.2,round-precision=2]
S[table-format=3.2,round-precision=2]
c
S[table-format=3.2,round-precision=2]
S[table-format=3.2,round-precision=2]}
\toprule
& \multicolumn{2}{c}{2018} && \multicolumn{2}{c}{2022} && \multicolumn{2}{c}{Difference} \\
\cmidrule{2-3} \cmidrule{5-6} \cmidrule{8-9}
\multicolumn{1}{l}{Variables} & {Mean} & {SD} && {Mean} & {SD} && {Mean} & {SE} \\
\midrule
Not at work & 0.325881&  0.468810  && 0.329113&  0.470022 && 0.003232 & 0.019266 \\
Age & 48.605632& 15.636059 && 49.235675& 15.632311 && 0.630042 & 0.579425 \\
Married, no kids and no others & 0.165682& 0.371880 && 0.159807& 0.366529 && -0.005876 & 0.011775 \\
Married, oldest child age $<$6 & 0.044030& 0.205210 && 0.033931& 0.181103 && -0.010099 &  0.008217 \\
Married, oldest child age 6--17 & 0.187978& 0.390784 && 0.186932& 0.389966 && -0.001046 & 0.018101 \\
Married, oldest child age $>$17 & 0.126248& 0.332205 && 0.126325&  0.332309 && 0.000077 & 0.015847 \\
Married, others & 0.073208& 0.260538 && 0.083697&  0.277011 && 0.010489 & 0.013611 \\
Single mother, oldest child age $<$18 & 0.076909& 0.266508 && 0.087964&    0.283322 && 0.011056  & 0.011198 \\
Single female & 0.101934& 0.302631 && 0.115491&  0.319703 && 0.013557\sym{*} & 0.007497 \\
None of the above family types & 0.224009& 0.417023 && 0.205852& 0.404436 && -0.018158 & 0.017738 \\
% No schooling or elementary & 0.032004& 0.176051 && 0.040477& 0.197130 && 0.008473 & 0.009473 \\
High school & 0.273085& 0.445646 && 0.271869& 0.445047 && -0.001216 & 0.018755 \\
Some college & 0.347369& 0.476243 && 0.324531& 0.468330 && -0.022838 & 0.019764 \\
Bachelor{'}s degree or above & 0.347542& 0.476298 && 0.363123& 0.481034 &&  0.015582  & 0.019496 \\
White & 0.616937&  0.486244 && 0.620049&  0.485510 && 0.003112 & 0.021162 \\
Black & 0.135737& 0.342587 && 0.116370& 0.320757 && -0.019367 &  0.015609 \\
Other races & 0.247326&  0.431556 && 0.263581& 0.440698  && 0.016256 & 0.019326 \\
Urban area & 0.932048& 0.251720 &&  0.948295&    0.221493  &&  0.016246\sym{*} & 0.009708 \\
\midrule
Number of observations & \multicolumn{2}{c}{2188} && \multicolumn{2}{c}{1788} && \multicolumn{2}{c}{} \\
\bottomrule
\end{tabular}
\begin{tablenotes}[para,flushleft]
\footnotesize{}
\item 
\textsuperscript{***}~$p<0.01$, \textsuperscript{**}~$p<0.05$, \textsuperscript{*}~$p<0.1$.
\end{tablenotes}
\end{threeparttable}
\end{table}

\Cref{tab:stat:female:1} presents the summary statistics for consumption and assets for the female-headed households. 
On average, female-headed households have lower consumption and asset holdings than male-headed households.
Female-headed households have experienced a 9.6\% ($=3.109/2.836-1$) increase in real consumption, comparable to the increase for male-headed households, but a much smaller 13.2\% ($=14.115/12.464-1$) increase in real assets from 2018 to 2022 compared to male-headed households. 
Female-headed households show a lower dispersion in log real assets than male-headed households in 2022, but greater dispersion in 2018. 
For the log real consumption, female-headed households are more concentrated than male-headed households in 2018 and are similar to or slightly more dispersed in 2022.

Unlike male-headed households that have observed many characteristic changes,
\cref{tab:stat:female:2} shows that most of the characteristics of female-headed households remain the same or have experienced insignificant little changes between the two years, except an increase in the share of single female households and an increase in the share who live in the urban area at a 10\% significance level.

The composition of female-headed households also differs from that of male-headed households.
Compared with male-headed households, female-headed households have a larger proportion of household heads who have not been working in the past 12 months.
The proportion of single mothers with the oldest child under 18 is 8 percent, whereas the proportion of single father counterpart is only 3 percent.
The single mother with the oldest kids, who are 18 or older, is counted in the ``none of the above'' category, which accounts for 22 percent, whereas the male-headed counterpart accounts for only 11 percent.
Other family types are similar across male-headed and female-headed households, 
There is a smaller proportion of female-headed household heads with a bachelor's degree or higher than in male-headed households.

\subsection{Empirical results for female-headed households}
\label{sec:female:emp}

\Cref{tab:emp:female:6factor} decomposes the consumption inequality changes for the female-headed households.

\begin{table}[htbp]
\centering
\caption{\label{tab:emp:female:6factor}Decomposition of consumption inequality changes for female-headed households}
\sisetup{round-precision=1,round-mode=places,detect-all}
\begin{threeparttable}
\begin{tabular}{lccccccc}
\toprule
& & \multicolumn{6}{c}{Effects of changes in} \\
\cmidrule{3-8}
{Statistics} 
  & {\makecell{Total\\Changes}} 
  & {Assets} 
  & {\makecell{College\\Education}}
  & {\makecell{Married\\No Kids}} 
  & {\makecell{Work\\Status}} 
  & {\makecell{Remaining\\Covariates}} 
  & {\makecell{Consumption\\Structure}} \\
\midrule

SD          
  & \num{0.631554}  (\num{1.524275})
  & \num{3.169980}  (\num{0.898017})
  & \num{0.642572}  (\num{0.328761})
  & \num{0.703771}  (\num{0.225882})
  & \num{0.435803}  (\num{0.243446})
  & \num{-5.454450} (\num{0.878969})
  & \num{1.133880}  (\num{1.655354}) \\[0.65em]
  
% Share estimates: Total=.; Assets=501.93286; Education=101.74444; Family=111.43475; Work=69.0048; Remaining=-863.65473; Structure=179.53788
% Share SEs: Total=.; Assets=393.01981; Education=69.500057; Family=84.273455; Work=47.019899; Remaining=664.99775; Structure=92.224501

90--10      
  & \num{-1.551817} (\num{5.745900})
  & \num{8.543926}  (\num{3.218659})
  & \num{1.933748}  (\num{1.314863})
  & \num{1.617576}  (\num{1.139597})
  & \num{1.600359}  (\num{0.808837})
  & \num{-16.824496} (\num{3.932090})
  & \num{1.577069}  (\num{5.403741}) \\[0.65em]
  
% Share estimates: Total=.; Assets=-550.57548; Education=-124.61186; Family=-104.23754; Work=-103.12806; Remaining=1084.1802; Structure=-101.62723
% Share SEs: Total=.; Assets=371.76138; Education=67.112759; Family=67.991722; Work=30.591524; Remaining=568.75125; Structure=104.13765

50--10      
  & \num{-2.316956} (\num{4.803413})
  & \num{5.349387}  (\num{2.578906})
  & \num{1.502183}  (\num{1.341477})
  & \num{1.031545}  (\num{0.990964})
  & \num{1.096713}  (\num{0.703489})
  & \num{-11.258092} (\num{3.759455})
  & \num{-0.038691} (\num{4.644354}) \\[0.65em]
  
% Share estimates: Total=.; Assets=-230.87998; Education=-64.834334; Family=-44.521556; Work=-47.334244; Remaining=485.9002; Structure=1.669918
% Share SEs: Total=.; Assets=221.26732; Education=40.554252; Family=28.250966; Work=16.887104; Remaining=396.47494; Structure=90.61732

90--50      
  & \num{0.765138}  (\num{3.022728})
  & \num{3.194539}  (\num{1.591652})
  & \num{0.431566}  (\num{0.449723})
  & \num{0.586032}  (\num{0.621482})
  & \num{0.503646}  (\num{0.465568})
  & \num{-5.566404} (\num{1.433676})
  & \num{1.615760}  (\num{2.825661}) \\[0.65em]
  
% Share estimates: Total=.; Assets=417.51132; Education=56.403612; Family=76.591581; Work=65.824126; Remaining=-727.50296; Structure=211.17232
% Share SEs: Total=.; Assets=250.61716; Education=23.401297; Family=71.529142; Work=19.211512; Remaining=431.57315; Structure=80.899263

75--25      
  & \num{0.143390} (\num{3.599661})
  & \num{3.516300} (\num{2.120152})
  & \num{1.489944} (\num{0.750874})
  & \num{2.085378} (\num{0.840657})
  & \num{0.159050} (\num{0.564738})
  & \num{-9.627883} (\num{2.268251})
  & \num{2.520601} (\num{3.300623}) \\[0.65em]
  
% Share estimates: Total=.; Assets=2452.2678; Education=1039.0872; Family=1454.3425; Work=110.92141; Remaining=-6714.4865; Structure=1757.8675
% Share SEs: Total=.; Assets=263.11297; Education=37.739966; Family=93.181804; Work=21.82765; Remaining=492.93191; Structure=99.100296

95--5       
  & \num{6.939634}  (\num{6.406917})
  & \num{12.774509} (\num{6.082820})
  & \num{0.996708}  (\num{2.610190})
  & \num{4.923851}  (\num{1.994884})
  & \num{4.147871}  (\num{2.053581})
  & \num{-21.414685} (\num{7.016351})
  & \num{5.511379}  (\num{7.480767}) \\[0.65em]
  
% Share estimates: Total=.; Assets=184.08046; Education=14.362552; Family=70.952604; Work=59.770751; Remaining=-308.58524; Structure=79.418877
% Share SEs: Total=.; Assets=160.74297; Education=41.727403; Family=36.978983; Work=32.932955; Remaining=258.00386; Structure=86.441827

Gini        
  & \num{0.003948} (\num{0.325214})
  & \num{0.630115} (\num{0.190086})
  & \num{0.134018} (\num{0.067543})
  & \num{0.157240} (\num{0.050653})
  & \num{0.090954} (\num{0.049617})
  & \num{-1.162297} (\num{0.184634})
  & \num{0.153918} (\num{0.339950}) \\
  % \\[0.65em]
  
% Share estimates: Total=.; Assets=15961.303; Education=3394.7704; Family=3983.0075; Work=2303.943; Remaining=-29441.875; Structure=3898.851
% Share SEs: Total=.; Assets=430.90301; Education=85.992167; Family=94.719037; Work=54.735339; Remaining=746.87992; Structure=106.0475

\bottomrule
\end{tabular}
\begin{tablenotes}[para,flushleft]
\footnotesize{}
% Female-headed households only.  
All values are in percentages.
Bootstrapped standard errors with 500 repetitions in parentheses.  
\end{tablenotes}
\end{threeparttable}
\end{table}

Unlike male-headed households, female-headed households exhibit a different pattern in the observed changes in consumption inequality measures between the two years.
The consumption inequality has increased slightly for the upper half of the distribution ($90{:}50$ ratio) and considerably for the larger range of the distribution ($95{:}5$ ratio).
This indicates that the poor female-headed households at the bottom of the distribution may benefit disproportionately less than the wealthy female-headed households.
For the lower half of the distribution (50:10 ratio), female-headed households have experienced a larger decline in the consumption inequality than male-headed households.

As with male-headed households, the rise in asset holdings among female-headed households has significantly increased consumption inequality across all measures. 
It helps poor households on the lower half of the distribution ($50{:}10$ ratio) less than those on the upper half ($90{:}50$ ratio).

Similar to male-headed households, changes in the remaining household characteristics for female-headed households have significantly reduced the consumption inequality in all measures. 
It reduces consumption inequality more on the lower half of the distribution ($50{:}10$ ratio) than on the upper half ($90{:}50$ ratio).

Unlike male-headed households, changes in the consumption structure for female-headed households have increased consumption inequality ratios, although statistically insignificant.

\end{document}